# Incentives for P2P-Assisted Content Distribution: If You Can't Beat 'Em, Join 'Em

Vinod Ramaswamy*, Sachin Adlakha†, Srinivas Shakkottai*, and Adam Wierman†

*Dept. of ECE, Texas A&M University, Email: {vinod83,sshakkot}@tamu.edu

† Department of Computing and Mathematical Sciences, Caltech, Email: {adlakha,adamw}@caltech.edu

*Abstract*—The rapid growth of content distribution on the Internet has brought with it proportional increases in the costs of distributing content. Adding to distribution costs is the fact that digital content is easily duplicable, and hence can be shared in an illicit peer-to-peer (P2P) manner that generates no revenue for the content provider. In this paper, we study whether the content provider can recover lost revenue through a more innovative approach to distribution. In particular, we evaluate the benefits of a hybrid revenue-sharing system that combines a legitimate P2P swarm and a centralized client-server approach. We show how the revenue recovered by the content provider using a server-supported legitimate P2P swarm can exceed that of the monopolistic scheme by an order of magnitude. Our analytical results are obtained in a fluid model, and supported by stochastic simulations.

## I. INTRODUCTION

THE past decade has seen the rapid increase of content distribution using the Internet as the medium of delivery [2]. Users and applications expect a low cost for content, but at the same time require high levels of quality of service. However, providing content distribution at a low cost is challenging. The major costs associated with meeting demand at a good quality of service are (i) the high cost of hosting services on the managed infrastructure of CDNs such as Akamai [3], [4], and (ii) the lost revenue associated with the fact that digital content is easily duplicable, and hence can be shared in an illicit peer-to-peer (P2P) manner that generates no revenue for the content provider. Together, these factors have led content distributors to search for methods of defraying costs.

One technique that is often suggested for defraying distribution costs is to use legal peer-to-peer (P2P) networks to supplement provider distribution [5]–[7]. It is well documented that the efficient use of P2P methods can result in significant cost reductions from the perspective of ISPs [3], [8]; however there are substantial drawbacks as well. Probably the most troublesome is that providers fear losing control of content ownership, in the sense that they are no longer in control of the distribution of the content and worry about feeding illegal P2P activity.

Thus, a key question that must be answered before we can expect mainstream utilization of P2P approaches is: *How can users that have obtained content legally be encouraged to reshare it legally?* Said in a different way, can mechanisms be designed that ensure legitimate P2P swarms will dominate the illicit P2P swarms?

In this paper, we investigate a "revenue sharing" approach to this issue. We suggest that users can be motivated to reshare the content legally by allowing them to share the revenue associated with future sales. This can be accomplished through either a lottery scheme or by simply sharing a fraction of the sale price. Recent work on using lotteries to promote societally beneficial conduct [9] suggests that such schemes could potentially see wide spread adoption.

Such an approach has two key benefits: First, obviously, this mechanism ensures that users are incentivized to join the legitimate P2P network since they can profit from joining. Second, less obviously, this approach actually damages the illicit P2P network. Specifically, despite the fact that content is free in the illicit P2P network, since most users expect a reasonable quality of service, if the delay in the illegitimate swarm is large they may be willing to use the legitimate P2P network instead. Thus, by encouraging users to reshare legitimately, we are averting them from joining the illicit P2P network, reducing its capacity and performance; thus making it less likely for others to use it.

The natural concern about a revenue sharing approach is that by sharing profits with users, the provider is losing revenue. However, the key insight provided by the results in this paper is that by discouraging users from joining illicit P2P network, the increased share (possibly exponentially more) of legitimate copies makes up for the cost of sharing revenue with end-users.

More specifically, the contribution of this paper is to develop and analyze a model to explore the revenue sharing approach described above. Our model (see Section II) is a fluid model that builds on work studying the capacity of P2P content distribution systems. The key novel component of the model is the competition for users among an illicit P2P system and a legal content distribution network (CDN), which may make use of a supplementary P2P network with revenue sharing. The main results of the paper (see Section III) are Theorems 1-4, which highlight the order-of-magnitude gains in revenue extracted by the provider as a result of participating in revenue sharing. Further, In addition to the analytic results, to validate the insights provided by our asymptotic analysis of the fluid model we also perform numerical experiments of the underlying finite stochastic model. Tables I and II summarize these experiments, which highlight both that the results obtained in the fluid model are quite predictive for the finite setting and that there are significant beneficial effects of revenue sharing.

There is a significant body of prior work modeling and analyzing P2P systems. Perhaps the most related work from this literature is the work that focuses on server-assisted P2P content distribution networks [10]–[15] in which a central server is used to "boost" P2P systems. This boost is important since pure P2P systems suffer poor performance during initial stages of content distribution. In fact, it is this initially poor performance that our revenue sharing mechanism exploits to ensure that the legitimate P2P network dominates.

Two key differentiating factors of the current work compared to this work are: (i) We model the impact of competition between legal and illegal swarms on the revenue extraction of a content provider. (ii) Unlike most previous works on P2P systems, we consider a time varying viral demand model for the evolution of demand in a piece of content based on the Bass diffusion model (see Section II). Thus, we model the fact that interest in content grows as interested users contact others and make them interested.

With respect to (i), there has been prior work that focuses on identifying the relative value of content and resources for different users [16], [17]. For instance, [16] deals with creating a content exchange that goes beyond traditional P2P barter schemes, while [17] attempts to characterize the relative value of peers in terms of their impact on system performance as a function of time. However, to the best of our knowledge, ours is the first work that considers the question of economics and incentives in hybrid P2P content distribution networks.

With respect to (ii), there has been prior work that considers fluid models of P2P systems such as [18]–[20]. However, these all focus on the performance evaluation of a P2P system with constant demand rate. As mentioned above, a unique facet of our approach is that we explicitly make use the transient nature of demand in our modeling. In the sense of explicitly accounting for transient demand, the closest work to ours is [14]. However, [14] focuses only on jointly optimizing server and P2P usage in the case of transient demand in order to obtain a target delay guarantee at the lowest possible server cost.

The remainder of the paper is organized as follows. We first introduce the details of our model in Section II. Then, Section III summarizes analytic and numeric results, the proofs of which are included in the appendix. Finally, Section V provides concluding remarks.

## II. MODEL OVERVIEW

Our goal is to model the competition between illicit peer-to-peer (P2P) distribution and a legitimate content distribution network (CDN), which may make use of its own P2P network. Our model is a fluid model, and there are four main components:

1) The evolution of the demand for content. A key feature of this paper is that we consider a realistic model for the evolution of demand, specifically, the Bass diffusion model.
2) The model of user behavior, which allows the user to strategically choose between attaining content legally or illegally based on the price and performance of the two options.

3) The model of the illicit P2P system.
4) The model of the legal CDN and its possibility to use "revenue sharing".

We discuss these each in turn in the following.

### A. The evolution of demand

The simplest possible model of demand is that the entire population gets interested in the content simultaneously at time $t = 0$. We call this the "Flash crowd model" due to the instantaneous appearance of all the demand. While the model is simplistic, it can serve as a foundation for developing performance results, and we will utilize it as our base case. More complex models of demand can be considered as well. Indeed, models of the dynamics of demand growth for innovations dates to the work of Griliches [21] and Bass [22]. The most widely used model for dynamics of demand growth is the Bass diffusion model which describes how new products get adopted as potential users interact with users that have already adopted the product. Such word of mouth interaction between users and potential users is very common in the Internet and we use a version of Bass diffusion model that only has word of mouth spreading. We describe both models formally below.

We define $N$ to be the total size of the population and $I(t)$ to be the number of users that are interested in the content at time $t$. In the Flash Crowd Model,

$$I(t) = N, \qquad (1)$$

since all users are interested from the very beginning. In the Bass diffusion model, each interested user "attempts" to cause a randomly selected user to become interested in the content.[1] At any time $t$, there are $N - I(t)$ users that could potentially be interested in the content. Thus, the probability of finding such a users is $(N - I(t))/N$. Assuming that an interested user can interact with other users at rate 1 per unit time, we get that the rate at which interested users increase is given by the following differential equation:

$$\frac{dI(t)}{dt} = \left( \frac{N - I(t)}{N} \right) I(t). \qquad (2)$$

The above differential equation can be easily solved and yields the so-called *logistic function* as its solution.

$$I(t) = \frac{I(0)e^t}{1 - (1 - e^t)\frac{I(0)}{N}}, \qquad (3)$$

where $I(0)$ is the number of user that are interested in the content at time $t = 0$.

Though the Bass model is quite simple, it is a useful qualitative summary of the spread of content. To highlight this, Figure 1 (taken from [14]) highlights a similar behavior in a data trace from CoralCDN [23], a CDN hosted at different university sites. The figure shows the cumulative demand for a home video of the Asian Tsunami seen over a month in December 2005. For comparision, the figure on the right shows the model in equation (3). The qualitative usefulness of the Bass model has been verified empirically in many settings, and hence the Bass model is often considered as canonical [24].

---

[1]Note that these "attempts" should not be interpreted literally, but rather as the natural diffusion of interest in the new content through the population.

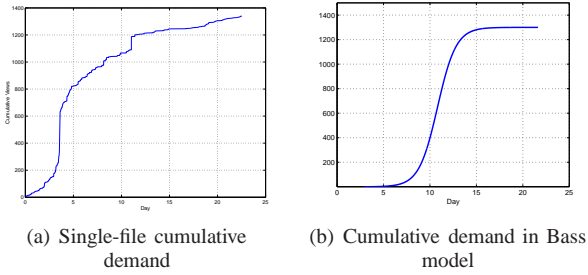

(a) Single-file cumulative demand

(b) Cumulative demand in Bass model

Fig. 1. (a) shows the cumulative demand for a file over one month on Coral CDN (Dec 2005–Jan 2006). (b) shows the cumulative demand seen in a Bass diffusion.

### B. The progression of a user

In order to capture the strategic behavior of users in the face of competition between a legitimate CDN using P2P and an illicit P2P network our model is necessarily complex. Figure 2 provides a broad overview of the user behavior in the system, which we explain in detail in the following.

Let us explain the model through tracking the progression of a user. We term an initial user that wants, but has not yet attained, the content a *Wanter (W)*. When a Wanter arrives to the system, it has two options: get content from the illicit P2P system for free or get content from the legitimate system for a price $p$. We assume that the Wanter wishes to obtain content as quickly and cheaply as possible, and so she first approaches the illicit P2P swarm and then only attains the content from the legitimate system if the content is not attained a reasonable time interval (one infinitesimal clock tick in our model) from the illicit P2P. This cycle repeats, if necessary, until the content is attained. In some sense, this is the worst-case for the legitimate provider since the illicit source is tried first.

Once the Wanter has attained the content (legally or illegally), it could stay in the system and assist in content dissemination. We denote the probability of this event by $\kappa < 1$. Otherwise, it could simply *Quit (Q)* and leave the system with probability $1 - \kappa$. Now, if a Wanter obtains the content *legally* and decides to assist in dissemination, it has two options: (i) It might decide to use the content to assist the illicit P2P swarm, i.e., go *Rogue (R)*. We denote the probability this happens by $\rho < 1$. (ii) It might decide to assist the legitimate P2P swarm (if one exists) as a *Booster (B)*. We denote the probability of this event by $\beta < 1$. Note that $\beta = 0$ if no legal P2P is used. Clearly $\rho + \beta = \kappa$. However, if a Wanter obtains content *illegally* and chooses to stay in the system, it can only aid the illicit swarm as a *Fraudster (F)*. The probability of this event is simply $\kappa$.

Note that the goal of revenue sharing is to incentivize Wanters to become Boosters after attaining content legally, rather than going Rogue. The hope is that the revenue invested toward reducing the number of "early adopters" that go Rogue keeps the illicit P2P swarm from growing enough to provide good enough quality of service to dominate the legitimate swarm.

To model this system more formally, we introduce the following notation. Let $N_w(t)$ be the number of Wanters at time $t$, i.e., the number of users who have not yet attained the content, and assume $N_w(0) = 0$. Further, let $N_l(t)$ and $N_i(t)$ be the number of users with legal and illegal copies of the content at time $t$. Note that the total number of interested users at any time $t$ satisfies the following equation

$$I(t) = N_w(t) + N_l(t) + N_i(t) \tag{4}$$

We can break this down further by noting that the number of Rogues, Fraudsters, and Boosters in the system at time $t$ (denoted by $N_r(t)$, $N_f(t)$, and $N_b(t)$ respectively) is:

$$N_r(t) = \rho N_l(t) \tag{5}$$
$$N_f(t) = \kappa N_i(t) \tag{6}$$
$$N_b(t) = \beta N_l(t), \tag{7}$$

with $\rho + \beta < 1$. The rest of legal and illegal users leave the system.

The key remaining piece of the model is to formally define the transition of Wanters to holders of illegal/legal content, i.e., the evolution of $N_i(t)$ and $N_l(t)$. However, this evolution depends critically on the model of the two systems, and so we describe it in the next section.

### C. System models

We discuss in detail the illicit and legitimate system models below. The factors in these models are key determinants of the choice of a Wanter to get the content legally or illegally. When modeling the two systems, we consider a fluid model, and so the performance is determined primarily by the capacity of each system, i.e., the combination of the initial seeds and the Fraudsters/Boosters that choose to join (and add capacity). However, other factors also play a role, as we describe below. Throughout, we model the upload capacity of a user as being one.

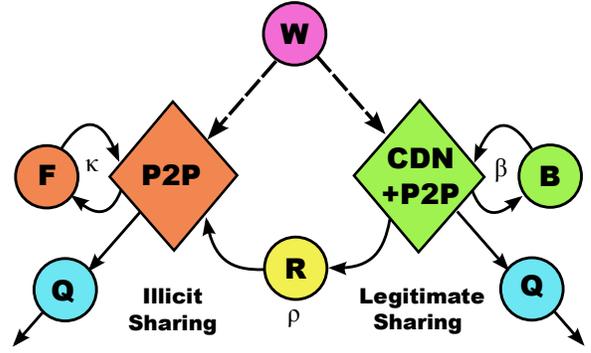

Fig. 2. An overview of the progression of a user through the systems. The labels are defined as follows: W - Wanter, F - Fraudster, R - Rogue, B - Booster, and Q - Quit.

*1) The illicit P2P system:* There are two components to the model of the illicit P2P network: (i) the efficiency of the network in terms of finding content, and (ii) the initial size of the network and its growth.

Let us start with (i). To capture the efficiency of the P2P system, we take a simple qualitative model. When attaining the content illegally, a Wanter must contact either a Rogue or a Fraudster. We let $\eta(t)$ capture the probability of a Wanter finding a Rogue or a Fraudster when looking for one instantaneous time slot. We consider two cases: an efficient P2P and an inefficient P2P. In an ***efficient P2P***, we model

$$\eta(t) = 1,$$

with the understanding the the P2P allows easy lookup of content and all content is truthfully represented. In contrast, for an ***inefficient P2P***, we model

$$\eta(t) = (N_r(t) + N_f(t))/N,$$

where recall that $N$ is the total population size. This corresponds to looking randomly within the user population for a Rogue or Fraudster. Neither of these models is completely realistic, but they provide lower and upper bounds to the true efficiency of an illicit P2P system.

Next, with respect to (ii), we model the initial condition for the illicit network with $N_i(0) = 0$, since the assumption is that the content has not yet been released, and therefore is not yet available in the illicit P2P swarm. From this initial condition, $N_i(0)$ evolves as follows:

$$\frac{dN_i(t)}{dt} = \min\left\{\eta(t)\left(N_w(t) + \frac{dI(t)}{dt}\right), N_r(t) + N_f(t)\right\},\tag{8}$$

The interpretation of the above is that $N_r(t) + N_f(t)$ is the current capacity of the illicit P2P and $\eta(t)\left(N_w(t) + \frac{dI(t)}{dt}\right)$ is the fraction of the Wanters (newly arriving and remaining in the system) that find the content in the illicit P2P network. The min operator then ensures that no more than the capacity is used.

*2) The legitimate CDN:* As discussed in the introduction, our goal in this work is to contrast the revenue attained by a CDN that uses P2P and revenue sharing with one that does not use P2P. Thus, there are two key factors in modeling the legitimate CDN: (i) the rate at which users that possess content copies become fraudsters or boosters, and (ii) the initial size of the CDN and its growth, which depends on the presence/absence of the legal P2P.

Let us start with (i). From a performance standpoint, the most important parameter is $\kappa$, since it determines what fraction of users stay in the system and act as servers. These users could either support the legal system as boosters, or the illegal one as fraudsters. The question that we wish to answer is that of how much of an impact the division of those who stay into fraudsters and boosters would have on revenue obtained. As we saw earlier,

$$\rho + \beta = \kappa,$$

and our key result will be on their relative impact on obtainable revenue. How we might attempt to control the booster factor $\beta$ through different amounts of revenue sharing requires further modeling of user motivation, which we will consider in greater detail in Section IV. But initially we are more concerned with the impact of $\rho$ and $\beta$, rather than how to socially engineer their values.

Next, with respect to (ii), unlike for the illicit P2P swarm, the legitimate network does not start empty. This is because it has a set of dedicated servers at the beginning which are then (possibly) supplemented using a P2P network. We denote by $C_N$ be the capacity of the dedicated CDN servers when the total population size is $N$. Note that this capacity must scale with the total population size to ensure that the average wait time for the users is small. As shown in [14], a natural scaling that ensures no more that $O(\ln \ln N)$ delay is to have the capacity $C_N = \Theta(N/\ln N)$. Based on this, we adopt

$$C_N = \frac{N}{\ln N}$$

in this work. Additionally, we assume $N_l(0) = 0$, in the case of Flash Crowd model and $N_l(0) = I(0)$ in the case of Bass model.

Given these initial conditions, $N_l(t)$ evolves as follows:

$$\frac{dN_l(t)}{dt} = \begin{cases} C_N + \beta N_l(t), & N_w(t) > 0, \\ \min\left\{C_N + \beta N_l(t), \frac{dI(t)}{dt} - \frac{dN_i(t)}{dt}\right\} & N_w(t) = 0. \end{cases}\tag{9}$$

The interpretation for the above is that if there are a positive number of Wanters remaining in the system, then the full current capacity of the CDN can be used to serve them, i.e., $C_N + \beta N_l(t)$. However, if there are no "leftover" Wanters, arriving Wanters that are not served by the illicit P2P ($\frac{dI(t)}{dt} - \frac{dN_i(t)}{dt}$) are served up to the capacity of the CDN.

## III. RESULTS

To characterize the performance of the CDN against the illicit P2P distribution, we use *fractional legitimate copies*, which is defined as follows:

**Definition 1.** *The **fractional legitimate copies**, $L$, is defined as*

$$L = \frac{N_l(T_\infty)}{N},\tag{10}$$

*where $T_\infty$ is defined as the time after which only $\Omega(\ln N)$ users are left in the system without a copy of the content*

Using this metric, we look at the performance of the CDN in two settings: when the CDN competes against inefficient illicit P2P sharing and when it competes against efficient illicit P2P sharing. Recall, that our models for these two cases are meant to serve as upper and lower bounds on the true efficiency of an illicit P2P system. We start by considering the case of an inefficient, illicit P2P. Note that the theorems stated below characterize only the asymptotic growth of the fractional legitimate copies. However, the proofs of these theorems, presented in Appendices A-D, actually characterize the exact growth.

## A. Inefficient illicit P2P

As discussed before, we look at the performance of CDN, under two simple models of demand evolutions, namely Flash Crowd Model (1) and Bass model (3).

First, we state the result for Flash Crowd model.

**Theorem 1.** *Suppose $I(t)$ satisfies* (1). *The fractional legitimate copies attained by the content provider in the presence an inefficient, illicit P2P is*

$$L \in \Omega\left(\frac{\ln\ln N + (\ln N)^{\frac{\beta}{\kappa}}}{\ln N}\right). \tag{11}$$

*Further, when $\beta = 0$,*

$$L \in \Theta\left(\frac{\ln\ln N}{\ln N}\right). \tag{12}$$

The interpretation of this theorem is striking. When booster factor, $\beta$, is zero, the fractional legitimate copies is exponentially small, $\Theta\left(\frac{\ln\ln N}{\ln N}\right)$. However, as $\beta$ increases, the fractional legitimate copies grows by orders of magnitude.

Now, we consider the second model for demand evolution, Bass model. For analytic reasons, we are not able to work with the exact Bass model. Thus, we approximate the logistic curve, (3), as follows:

$$I(t) = \begin{cases} \frac{NI(0)e^t}{N - I(0) + I(0)e^t} & 0 \le t \le T_1 & : \text{Phase 1} \\ I_2 = N/\ln N & T_1 < t \le T_2 & : \text{Phase 2} \\ I_3 = \frac{N}{2} & T_2 < t \le T_3 & : \text{Phase 3} \\ I_4 = N & T_3 < t < T_4 & : \text{Phase 4} \end{cases} \tag{13}$$

where we have $T_1 = \ln(N/(I(0)\ln N))$, $T_2 = \ln(N/I(0))$, $T_3 = 2\ln(N/I(0))$ and $T_4 = 3\ln(N/I(0))$.[2] Notice that the first stage is the exact Bass diffusion, while the other stages are order sense approximations of the actual expression. Though this model is approximate, it yields the same qualitative insight as the original model. Now, we are ready to state the result.

**Theorem 2.** *Suppose $I(t)$ satisfies* (13). *The fractional legitimate copies attained by the content provider in the presence an inefficient, illicit P2P is*

$$L \in \Omega\left(\frac{\ln\ln N + (\ln N)^{\frac{\beta}{\kappa}}}{\ln N}\right) \tag{14}$$

*Further, when $\beta = 0$,*

$$L \in \Theta\left(\frac{\ln\ln N}{\ln N}\right). \tag{15}$$

Note that the results of the above theorem match with that of Theorem 1. That means, the fractional legitimate copies attained by the CDN under Bass model of evolution is no different from that of Flash Crowd model in asymptotic sense.

Next, let us consider the case of an efficient, illicit P2P system.

---

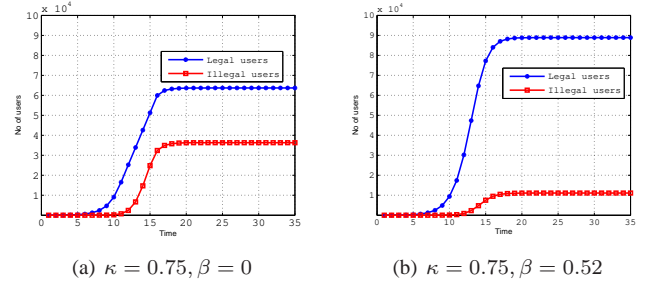

(a) $\kappa = 0.75, \beta = 0$     (b) $\kappa = 0.75, \beta = 0.52$

Fig. 3. Evolution of usage in the presence of inefficient illicit P2P sharing.

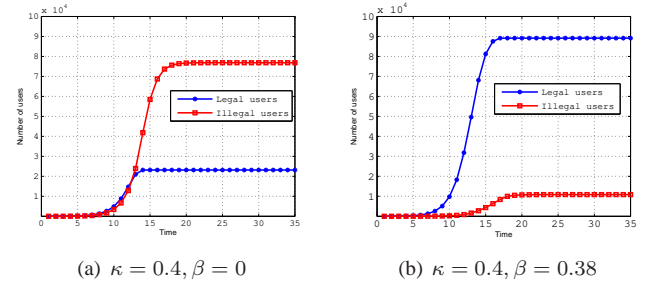

(a) $\kappa = 0.4, \beta = 0$     (b) $\kappa = 0.4, \beta = 0.38$

Fig. 4. Evolution of usage in the presence of efficient illicit P2P sharing.

## B. Efficient illicit P2P

As before, we first consider the case of Flash Crowd model.

**Theorem 3.** *Suppose $I(t)$ satisfies* (1). *Let $\kappa \in (0, 1 - I(0)/N)$. The fractional legitimate copies attained by the content provider in the presence an efficient, illicit P2P is*

$$L \in \Omega\left(\frac{1}{\ln N}\frac{(\ln N)^{\frac{\beta}{\kappa}} - 1}{\left(\frac{\beta}{\kappa}\right)}\right). \tag{16}$$

*Further, when $\beta = 0$,*

$$L \in \Theta\left(\frac{\ln\ln N}{\ln N}\right). \tag{17}$$

Again, the fractional legitimate copies rises by an order of magnitude as the booster factor, $\beta$, increases. Interestingly, the efficiency of the illicit P2P does not impact the asymptotic order of the fractional revenue when $\beta = 0$, since in both the efficient and inefficient case it is $\Theta\left(\frac{\ln\ln N}{\ln N}\right)$. However, the efficiency of the illicit P2P does affect the fractional legitimate copies attained for positive values of booster factor. In particular, it causes a $(1 - \frac{\beta}{\kappa})$ factor change in the fractional legitimate copies attained; however this has almost no effect on the asymptotic growth.

Now, we consider the second case, Bass model of evolution.

**Theorem 4.** *Suppose $I(t)$ satisfies* (3). *Let $\kappa \in (0, 1 - I(0)/N)$. The fractional legitimate copies attained by the content provider in the presence an efficient, illicit P2P is*

$$L \in \Omega\left(\frac{1}{\ln N}\frac{(\ln N)^{\frac{\beta}{\kappa}} - 1}{\left(\frac{\beta}{\kappa}\right)}\right). \tag{18}$$

*Further, when $\beta = 0$,*

$$L \in \Theta\left(\frac{\ln\ln N}{\ln N}\right). \tag{19}$$

The above theorem along with Theorem 3 asserts that the fractional legitimate copies attained by the CDN under Bass model of evolution is no different from that of Flash Crowd model in asymptotic order.

Since Theorems 1 and 3 rely on a fluid model, and characterize only the asymptotic growth rate of the fractional legitimate copies produced in the system, we present numerical simulations to verify the qualitative insights in discrete systems with finite $N$.

To simulate the underlying discrete stochastic system, we assume time is discrete and that there are $N = 100,000$ users in the system. A Bass model based interest evolution is assumed. That means, at each time slot, each user picks a Poisson distributed number (with mean 1) of other users to spread interest to. The server has a FIFO policy with service rate $C = 8000 \approx N / \ln N$.

Figure 3 illustrates the evolution of legal and illegal copies of the content in the case of an inefficient illicit P2P system with $\kappa = 0.75$. In Figure 3(a), where $\beta = 0$, the final number of legal copies produced in the system is $63,000$. When the booster factor increases, as shown in Figure 3(b) where $\beta = 0.52$, the number of legal copies increases to $88,888$; In fact, the fractional legitimate copies increases by more than $25\%$.

| $\frac{\beta}{\kappa}$ | $\kappa = 0.75$ | | $\kappa = 0.5$ | |
|---|---|---|---|---|
| | SIM* | ANL** | SIM | ANL |
| 0 | 0.64 | 0.60 | 0.69 | 0.67 |
| 0.10 | 0.71 | 0.71 | 0.77 | 0.75 |
| 0.24 | 0.77 | 0.72 | 0.82 | 0.77 |
| 0.41 | 0.81 | 0.75 | 0.86 | 0.79 |
| 0.63 | 0.87 | 0.79 | 0.92 | 0.80 |
| 0.92 | 0.97 | 0.85 | 0.98 | 0.82 |

In Table I, we compare the simulation results (SIM column entries in Table I ) against our analytical results (ANL column entries in Table I) from Lemma 9 and Corollary 10, for various combinations of $\kappa$ and $\beta$. As expected from Corollary 10, our analytical predictions closely match with the simulation results in the case, $\beta = 0$. In the case, $\beta > 0$, the predicted values are less than those obtained using simulation, which agrees with Lemma 9; nevertheless, the differences are quite small. Also observe that, as $\beta$ increases, the fractional legitimate copies improves significantly. Especially, in the case, $\kappa = 0.75$, as booster factor increases from $\beta = 0$ to $\beta = 0.92\kappa$, the fractional legitimate copies increases by $150\%$.

Next, we move to the case of an efficient illicit P2P. Figure 4 illustrates the case of an efficient illicit P2P system. In Figure 4(a), where $\beta = 0$, the final number of legal copies produced in the system is $45,920$. When the booster factor increases, as shown in Figure 4(b) where $\beta = 0.38$, the number of legal copies increases to $96,380$; In fact, the fractional legitimate copies increases by more than $100\%$.

In Table II, we tabulate the simulation results and the analytical results. The analytical results are obtained from Lemma 13 and Lemma 14. The simulation results are in agreement with our analytical predictions. Also note that, the improvement attained in the fractional legitimate copies, as $\beta$

| $\frac{\beta}{\kappa}$ | $\kappa = 0.75$ | | $\kappa = 0.5$ | | $\kappa = 0.25$ | |
|---|---|---|---|---|---|---|
| | SIM* | ANL** | SIM | ANL | SIM | ANL |
| 0 | 0.03 | 0.03 | 0.15 | 0.15 | 0.42 | 0.37 |
| 0.48 | 0.07 | 0.07 | 0.28 | 0.26 | 0.56 | 0.50 |
| 0.69 | 0.18 | 0.14 | 0.40 | 0.38 | 0.67 | 0.59 |
| 0.84 | 0.30 | 0.24 | 0.54 | 0.52 | 0.77 | 0.68 |
| 0.95 | 0.55 | 0.41 | 0.78 | 0.69 | 0.9 | 0.78 |

increase, is phenomenal. For example, in the case, $\kappa = 0.75$, as booster factor increases from $\beta = 0$ to $\beta = 0.95\kappa$, the fractional legitimate copies increases by $1833\%$.

## IV. Revenue Sharing Model

In the previous sections, we studied the impact of the three parameters $\rho$, $\beta$ and $\kappa$ on the eventual number of legal content copies in the system. We made the assumption that $\rho + \beta = \kappa$, following the intuition that $\kappa$ is the fixed probability of a user who has the content being willing to redistribute it, and *which* P2P swarm is joined affects the number of legal copies. We now consider the motivation behind the users' decisions on which swarm to join.

Suppose that the purchase price of a copy of the content is $p$. Hence, a user that wishes to obtain a legal copy of the content must pay the content generator the sum $p$ through some kind of online banking system. Suppose that the content owner utilizes a simple model for revenue sharing, where a user receives $\epsilon p$ for each piece of content it distributes when taking part in the legitimate network as a Booster. Thus, $\epsilon = 0$ corresponds to no revenue sharing. Note that this could potentially be implemented on a system such as BitTorrent by simply keeping track of amount uploaded by each peer[3]. The value $\epsilon$ can be viewed either as a share of the revenue from each download or as the expected payoff of a lottery scheme operated by the CDN.

While it is difficult to exactly predict the effect of revenue sharing, it seems reasonable that increased revenue sharing should limit the likelihood of a Wanter going rogue after attaining the content legally. To qualitatively capture this effect, we model $\rho$ as a decreasing function of $\epsilon$. A specific form could be

$$\rho = \kappa \phi(\epsilon),$$

where $\phi(.)$ is a decreasing function with $\phi(0) = 1$ and $\phi(1) = 0$.

Recall that we defined the parameter $R$ as the fractional revenue, also the fraction of legitimate copies in the system at $T_\infty$. It is clear that the profit obtained by the content owner also depends on the amount of revenue shared with the boosters, which in turn depends on the exact form of $\phi(\epsilon)$. Hence, the content owner would have to determine the optimal amount of revenue sharing in order to maximize profit. For illustration, let us choose

$$\phi(\epsilon) = N^{-\epsilon},$$

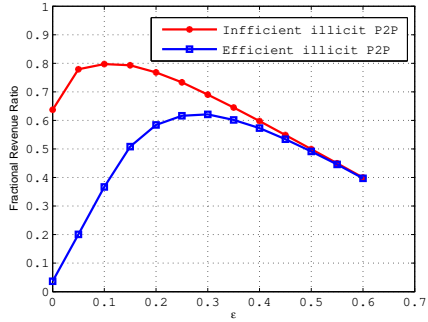

Fig. 5. Impact of the amount of revenue sharing on the fractional revenue attained by the CDN.

in our simulations. The results are shown in Figure 5, which illustrates the impact of the amount of revenue sharing on the fractional revenue ratio of the CDN in the cases of inefficient and efficient illicit P2Ps. We use $\kappa = 0.75$ in the simulation. The key point to observe in the figure is that there is a clear optimal amount of revenue sharing for the provider. In both cases, this amount is fairly small, however, it is clearly desirable to share more revenue in the presence of an efficient illicit P2P than in the presence of an inefficient illicit P2P. In fact, sharing nearly zero percent of the revenue still provides fairly close to the optimal fractional revenue in the inefficient case, while one must share more than $10\%$ of the revenue to be near-optimal in the case of an efficient, illicit P2P.

## V. CONCLUSION

Our goal in this paper is to quantify the ramifications of coopting legal P2P content sharing, not only as a means of reducing costs of content distribution, but more importantly, as a way of hurting the performance of illegal P2P file sharing. The model that we propose internalizes the idea that demand for any content is transient, and that all content will eventually be available for free through illegal file sharing. The objective then is not to cling to ownership rights, but to extract as much revenue from legal copies as possible within the available time. We develop a revenue sharing scheme that recognizes the importance of early adopters in extending the duration of time that revenue may be extracted. In particular, keeping users from "going rogue" (becoming seeds in illegal networks) by allowing them to extract some revenue for themselves (and so defray part of their expense in purchasing the content in the first place), provides *order sense improvements* in the extractable revenue. We realize that our paradigm is contrary to the "conventional wisdom" of charging *more* rather than *less* to early adopters, and also to discourage file sharing using legal threats. However, as many recent studies have demonstrated, incentives work better than threats in human society, and adoption of our revenue sharing approach might result in a cooperative equilibrium between content owners, distributors and end-users. Future work includes a characterization of the exact value of users based on their times of joining the system, as well as considering content streaming, which requires strict quality of service guarantees.

## REFERENCES

[1] V. Ramaswamy, S. Adlakha, S. Shakkottai, and A. Wierman, "Incentives for P2P-Assisted Content Distribution: If You Can't Beat 'Em, Join 'Em," in *Proceedings of Allerton Conference*, October 2012.

[2] C. Labovitz, D. McPherson, and S. Iekel-Johnson, "2009 Internet Observatory report," in *NANOG-47*, October 2009.

[3] W. B. Norton, "Internet Video: The Next Wave of Massive Disruption to the U.S. Peering Ecosystem," 2007, Available at http://www.equinix.com.

[4] "Akamai Technologies," http://www.akamai.com, 2011.

[5] "Pando Networks, Inc." http://www.pando.com/, 2011.

[6] "Rawflow, Inc." http://www.rawflow.com/, 2011.

[7] P. Rodriguez, S. Tan, and C. Gkantsidis, "On the feasibility of commercial, legal P2P content distribution," *ACM SIGCOMM Computer Communication Review*, vol. 36, no. 1, pp. 75–78, 2006.

[8] T. Karagiannis, P. Rodriguez, and K. Papagiannaki, "Should Internet service providers fear peer-assisted content distribution?" in *Proc. of the 5th ACM SIGCOMM conference on Internet Measurement*, 2005.

[9] D. Merugu, B. Prabhakar, and N. Rama, "An incentive mechanism for decongesting the roads: A pilot program in Bangalore," in *Proc. NetEcon, ACM Workshop on the Economics of Networked Systems*, July 2009.

[10] J. Wang, C. Yeo, V. Prabhakaran, and K. Ramchandran, "On the role of helpers in peer-to-peer file download systems: Design, analysis and simulation," in *Proc. IPTPS*, Feb. 2007.

[11] E. Setton and J. Apostolopoulos, "Towards quality of service for peer-to-peer video multicast," in *Proc. ICIP*, Sep. 2007.

[12] S. Liu, R. Zhang-Shen, W. Jiang, J. Rexford, and M. Chiang, "Performance bounds for peer-assisted live streaming," in *Proc. ACM SIGMETRICS*, Jun. 2008.

[13] M. Chen, M. Ponec, S. Sengupta, J. Li, and P. A. Chou, "Utility maximization in peer-to-peer Systems," in *Proc. ACM SIGMETRICS*, Jun. 2008.

[14] S. Shakkottai and R. Johari, "Demand aware content distribution on the Internet," *IEEE/ACM Transactions on Networking*, vol. 18, no. 2, April 2010.

[15] P. Parag, S. Shakkottai, and I. Menache, "Service routing in multi-ISP peer-to-peer content distribution: Local or remote?" in *Proc. of GameNets*, 2011.

[16] C. Aperjis and R. Johari, "A peer-to-peer system as an exchange economy," in *Proc. GameNets*, Oct. 2006.

[17] V. Misra, S. Ioannidis, A. Chaintreau, and L. Massoulié, "Incentivizing peer-assisted services: A fluid Shapley value approach," in *ACM SIGMETRICS Performance Evaluation Review*, vol. 38, no. 1, 2010, pp. 215–226.

[18] D. Qiu and R. Srikant, "Modeling and performance analysis of BitTorrent-like peer-to-peer networks," in *Proc. ACM SIGCOMM*, Aug. 2004.

[19] L. Massoulie and M. Vojnovic, "Coupon replication systems," in *Proc. ACM SIGMETRICS*, Jun. 2005.

[20] X. Yang and G. de Veciana, "Performance of Peer-to-Peer Networks: Service Capacity and Role of Resource Sharing Policies," *Performance Evaluation: Special Issue on Performance Modeling and Evaluation of P2P Computing Systems*, vol. 63, 2006.

[21] A. Griliches, "Hybrid Corn and the Economics of Innovation," *Science*, vol. 132, pp. 275–280, 1960.

[22] F. M. Bass, "A new product growth model for consumer durables," *Management Science*, vol. 15, pp. 215–227, 1969.

[23] M. J. Freedman, E. Freudenthal, and D. Mazières, "Democratizing content publication with Coral," in *Proc. NSDI*, Mar. 2004.

[24] G. Moore, *Crossing the Chasm: Marketing and Selling High-Tech Products to Mainstream Customers, Rev edition*. New York, NY: HarperBusiness, 1999.

## APPENDIX A
## PROOF OF THEOREM 1

To prove Theorem 1, we analyze two processes $\bar{N}_l(t)$ and $\bar{N}_i(t)$ which bounds the actual evolutions $N_l(t)$ and $N_i(t)$. Importantly, the bounding processes are equivalent to the original processes when $\beta = 0$.

Before stating the results, we introduce a few notation. Let

$$\theta_1 = \frac{\kappa}{2} + \frac{\kappa}{2}\sqrt{1 + \frac{4}{\kappa \ln N}}, \quad \theta_2 = \frac{\kappa}{2} - \frac{\kappa}{2}\sqrt{1 + \frac{4}{\kappa \ln N}},$$

$$b = -\frac{\theta_1}{\theta_2}, \quad \Delta\theta = \theta_1 - \theta_2, \tag{20}$$

$$\bar{\tau} = \frac{2}{\Delta\theta}\ln\left(\frac{\sqrt{1 + \frac{4}{\kappa \ln N}} + 1}{\sqrt{1 + \frac{4}{\kappa \ln N}} - 1}\right), \tag{21}$$

$$\bar{N}_l = \frac{\kappa C_N}{\beta\theta_1}\left(\frac{1}{1+b}\right)^{\frac{\beta}{\kappa}}\left(1 - e^{\left(-\frac{\beta\theta_1\bar{\tau}}{2\kappa}\right)}\right)e^{\left(\frac{\beta\theta_1}{\kappa}\bar{\tau}\right)}$$

$$- \frac{\kappa C_N}{\beta\theta_2}\left(\frac{1}{1+b}\right)^{\frac{\beta}{\kappa}}e^{\left(\frac{\bar{\tau}\beta}{2}\right)}\left(1 - e^{\frac{\beta\theta_2\bar{\tau}}{2\kappa}}\right). \tag{22}$$

Finally, we are ready to define the bounding processes used in the proof, $\bar{N}_l(t)$ and $\bar{N}_i(t)$. Let $\bar{N}_i(0) = N_i(0)$. Furthermore, let

$$\frac{d\bar{N}_i(t)}{dt} = \frac{\rho\bar{N}_l(t) + \kappa\bar{N}_i(t)}{N}(N - (\bar{N}_l(t) + \bar{N}_i(t))). \tag{23}$$

Similarly, let $\bar{N}_l(0) = N_l(0)$ and

$$\frac{d\bar{N}_l(t)}{dt} = \begin{cases} C_N + \beta\bar{N}_l(t)\frac{N - (\bar{N}_l(t) + \bar{N}_i(t))}{N}, & \bar{N}_w(t) > 0. \\ 0, & \bar{N}_w(t) = 0. \end{cases} \tag{24}$$

where $\bar{N}_w(t) = N - (\bar{N}_i(t) + \bar{N}_l(t))$.

We can now state our result characterizing the number of legal and illegal copies.

**Lemma 5.** *In the presence of an inefficient, illicit P2P, the number of illegal and legal copies at the end of evolution is*

$$N_l(T_\infty) \geq \bar{N}_l,$$

*where equality holds when $\beta = 0$.*

*Proof:* Recall that the efficiency factor of an inefficient illicit P2P, $\eta(t)$, is given by

$$\eta(t) = \frac{N_r(t) + N_f(t)}{N} = \frac{\rho N_l(t) + \kappa N_i(t)}{N}. \tag{25}$$

The second equality follows from (5) and (6). From (8), the illegal growth rate is

$$\frac{dN_i(t)}{dt} \stackrel{(a)}{=} \eta(t)N_w(t) \tag{26}$$

$$\stackrel{(b)}{=} \frac{(\rho N_l(t) + \kappa N_i(t))(N - (N_l(t) + N_i(t)))}{N}. \tag{27}$$

(a) follows from the definition of $\eta(t)$ and the fact that $N_w(t) \leq N$. (b) follows from (25) and (4). From equation (9), the growth rate of legal copies is given by

$$\frac{dN_l(t)}{dt} = \begin{cases} C_N + \beta N_l(t), & N_w(t) > 0, \\ 0, & N_w(t) = 0. \end{cases} \tag{28}$$

Let $U(t)$ be the total copies of the content in the system. Then, $U(t) = N_l(t) + N_i(t)$.

Now, we claim that,

$$N_l(T_\infty) \geq \bar{N}_l(T_\infty), \tag{29}$$

and the equality holds when $\beta = 0$.

The proof is as follows: First, we define, $\bar{U}(t) = \bar{N}_l(t) + \bar{N}_i(t)$. We can obtain $\frac{dN_i}{dU}$ and $\frac{d\bar{N}_i}{d\bar{U}}$ from the pair of equations (26), (28) and (23), (24) respectively. Then, it can be shown that

$$\frac{dN_i}{dU}|_{N_i=x, U=y} \leq \frac{d\bar{N}_i}{d\bar{U}}|_{\bar{N}_i=x, \bar{U}=y}, \tag{30}$$

and the equality holds when $\beta = 0$. Note that the range space of functions $U(t)$ and $\bar{U}(t)$ are identical. Since, the initial values $N_i(0)$ and $\bar{N}_i(0)$ are equal by definition, we get the result in (29).

Now, we derive $\bar{N}_l(t)$. Let $\bar{\tau}$ be the time at which the number of wanters in the system vanishes to zero. Then, $\bar{N}_w(t) = 0$ and $\bar{U}(t) = N$ for $t \in [\bar{\tau}, T_\infty]$. Adding (24) and (23), for $t \in (0, \bar{\tau}]$, we get,

$$\frac{d\bar{U}}{dt} = \left((\beta + \rho)\bar{N}_l(t) + \kappa\bar{N}_i(t)\right)\frac{(N - (\bar{N}_l(t) + \bar{N}_i(t)))}{N}$$

$$\stackrel{(f)}{=} \kappa\bar{U}(t)\frac{N - \bar{U}(t)}{N}.$$

(f) follows from the fact that $\rho + \beta = \kappa$ and the definition of $\bar{U}(t)$.

The above differential equation is in the form of a standard Riccatti equation, and it's solution can be written as

$$\bar{U}(t) = \frac{N\theta_2}{\kappa} + \frac{N\Delta\theta/\kappa}{1 + be^{-\Delta\theta t}}, \tag{31}$$

where $\Delta\theta = \theta_1 - \theta_2$. $\theta_1, \theta_2$ and $b$ are given by equation (20). From the relation, $\bar{U}(\bar{\tau}) = N$, we get (21).

Now, from (24), for $t \in (0, \bar{\tau}]$, we get

$$\frac{d\bar{N}_l(t)}{dt} = C_N + \beta\bar{N}_l(t)\frac{N - (\bar{N}_l(t) + \bar{N}_i(t))}{N}.$$

A lower bound on the solution of the above differential equation is provided by Lemma 16 in Appendix E. From the definitions of $b$ and $\bar{\tau}$, given by (20) and (21), it is clear that $b > 1$ and $\bar{\tau} > \ln b/\Delta\theta$. Then, by evaluating (147) at $t = \bar{\tau}$ with $\bar{N}_l(0) = I(0)$, we get $\bar{N}_l$ in (22). Also, when $\beta = 0$, the lemma yields an exact solution of the above differential equation. Hence proved. ∎

As mentioned in the statement of Lemma 5, the inequality is exact in the case of $\beta = 0$. Additionally, in this case, the form of $N_l(T_\infty)$ simplifies.

**Corollary 6.** *Let $\beta = 0$. In the presence of an inefficient, illicit P2P, the number of illegal and legal copies is given by*

$$N_l(T_\infty) = \frac{2C_N}{\Delta\theta}\ln\left(\frac{\sqrt{1 + \frac{4}{\kappa \ln N}} + 1}{\sqrt{1 + \frac{4}{\kappa \ln N}} - 1}\right). \tag{32}$$

Now that we have characterized the number of legal and illegal copies precisely, attaining the statement in Theorem 1 is accomplished by studying the asymptotics of the results in Lemma 5 and Corollary 6.

To begin, recall from (10) that,

$$L = \frac{N_l(T_\infty)}{N} \geq \frac{\bar{N}_l}{N}, \tag{33}$$

where $\bar{N}_l$ is defined by (22). Following a few algebraic steps, from the above equation, we get that

$$L \in \Omega\left(\frac{\ln\ln N + (\ln N)^{\frac{\beta}{\kappa}}}{\ln N}\right) \tag{34}$$

and $L \in \Theta\left(\frac{\ln\ln N}{\ln N}\right)$ if $\beta = 0$, which completes the proof of Theorem 1.

## Appendix B
## Proof of Theorem 2

To prove Theorem 2, we will go through a sequence of intermediate results characterizing the number of legal/illegal copies at the transition points of the approximate Bass model.

We start by characterizing the number of legal and illegal copies at the end of Phase 1.

**Lemma 7.** *In the presence of an inefficient, illicit P2P, the number of illegal and legal copies at the end of Phase 1 of the approximate Bass model are given by*

$$N_i(T_1) = \left(\frac{\rho I(0)}{\kappa - \rho} + \frac{N\rho}{(\kappa - \rho)^2}\right)\exp\left(B_N\right)$$
$$- \frac{I(T_1)\rho}{\kappa - \rho} - \frac{N\rho}{(\kappa - \rho)^2} \tag{35}$$

$$N_l(T_1) = I(T_1) - N_i(T_1), \tag{36}$$

*where*

$$I(T_1) = \frac{N}{\ln N}\frac{N}{N - I(0) + (N/\ln N)}$$
$$B_N = \left(\frac{(\kappa - \rho)}{N}(I(T_1) - I(0))\right).$$

Note that in the above, we have allowed $\kappa$, $\rho$, and $\beta$ to be arbitrary. In fact, in this case, $\beta$ is inconsequential since the full amount of interested copies can be served by the dedicated capacity of the CDN. Note that in the case when $\rho = \kappa$, things simplify considerably.

**Corollary 8.** *Let $\rho = \kappa$. In the presence of an inefficient, illicit P2P, the number of illegal and legal copies at the end of Phase 1 of the approximate Bass model are given by*

$$N_i(T_1) = \frac{\kappa(I^2(T_1) - I^2(0))}{2N}$$
$$N_l(T_1) = I(T_1) - N_i(T_1),$$

*where $I(T_1) = \frac{N}{\ln N}\frac{N}{N - I(0) + (N/\ln N)}$.*

We now prove the lemma.

*Proof of Lemma 7:* From equation (13), the population of interested copies in phase $I$ is given by

$$I(t) = \frac{NI(0)e^t}{N - I(0) + I(0)e^t}. \tag{37}$$

From the above equation, it is easy to verify that the rate of growth of interested copies is less than the server capacity $C_N$, i.e., $dI(t)/dt \leq C_N$. Thus, any interested user is served instantaneously either by a legal or illegal mechanism. Hence, the number of Wanters in the system is zero, i.e, $N_w(t) = 0$.

Therefore, it follows from equation (4) that $N_l(t) + N_i(t) = I(t)$.

Next, from equation (8), we get that

$$\frac{dN_i(t)}{dt} = \min\left\{\eta(t)\frac{dI(t)}{dt}, N_r(t) + N_f(t)\right\}$$
$$\overset{(a)}{=} \eta(t)\frac{dI(t)}{dt}, \tag{38}$$

where the equality (a) follows from the definition of $\eta(t)$ and the fact that $dI(t)/dt \leq C_N < N$.

Because we are considering an inefficient P2P, we have

$$\eta(t) = \frac{N_r(t) + N_f(t)}{N},$$
$$\overset{(b)}{=} \frac{\rho N_l(t) + \kappa N_i(t)}{N},$$
$$\overset{(c)}{=} \frac{\rho(I(t) - N_i(t))}{N} + \frac{\kappa N_i(t)}{N},$$
$$= \frac{\rho I(t)}{N} + \frac{(\kappa - \rho)N_i(t)}{N}.$$

where equality (b) follows from (5), (6) and the equality (c) follows from the fact that $N_l(t) = I(t) - N_i(t)$. Substituting the above result in equation (38), we get

$$\frac{dN_i(t)}{dt} = \frac{dI(t)}{dt}\frac{\rho I(t)}{N} + \frac{dI(t)}{dt}\frac{(\kappa - \rho)N_i(t)}{N}.$$

The solution of the above differential equation is given by

$$N_i(t) = K\exp\left(\frac{I(t)(\kappa - \rho)}{N}\right) - \frac{\rho I(t)}{\kappa - \rho} - \frac{N\rho}{(\kappa - \rho)^2},$$

where the constant $K$ can be obtained from the fact that $N_i(0) = 0$. Thus, the evolution of illegal copies is given by

$$N_i(t) = \left(\frac{\rho I(0)}{\kappa - \rho} + \frac{N\rho}{(\kappa - \rho)^2}\right)\exp\left(\frac{(\kappa - \rho)}{N}(I(t) - I(0))\right)$$
$$- \frac{\rho I(t)}{\kappa - \rho} - \frac{N\rho}{(\kappa - \rho)^2}.$$

The number of illegal copies at the end of Phase 1 can be obtained by evaluating the above expression at $t = T_1$. The remaining population get the content legally, i.e, $N_l(T_1) = I(T_1) - N_i(T_1)$. ∎

Now that we have characterized the number of legal and illegal copies at the end of Phase 1, we can move to Phases 2-4. Unfortunately, the resulting number of legal and illegal copies at the end of these phases is much more complicated. However, much of this complicated form is only necessary to specify the exact analytic values. Once we focus on the asymptotic form (as in Theorem 1), it simplifies considerably.

Before stating the result, we need to introduce a considerable amount of notation. This notation stems from the fact that we do not analyze the exact process of $N_l(t)$ and $N_i(t)$. Instead, we define a processes $\bar{N}_l(t)$ and $\bar{N}_i(t)$ which bounds $N_l(t)$ and $N_i(t)$ and analyze these processes. Importantly, the bounding processes are equivalent to the original processes when $\beta = 0$, i.e., the case of no revenue sharing. Before defining $\bar{N}_l$ and $\bar{N}_i$, Let

$$
\Delta\bar{\tau}_2 = \frac{1}{\kappa \ln N Z_1} \ln\left( \frac{Z_1 + 1 - \frac{2I(T_1)}{(N/\ln N)}}{Z_1 - 1 + \frac{2I(T_1)}{(N/\ln N)}} \right)
$$
$$
+ \frac{1}{\kappa \ln N Z_1} \ln\left( \frac{Z_1 + 1}{Z_1 - 1} \right), \tag{39}
$$
$$
\Delta\bar{\tau}_3 = \frac{2}{\kappa Z_2} \ln\left( \frac{Z_2 + 1 - \frac{4}{\ln N}}{Z_2 - 1 + \frac{4}{\ln N}} \right)
$$
$$
+ \frac{2}{\kappa Z_2} \ln\left( \frac{Z_2 + 1}{Z_2 - 1} \right), \tag{40}
$$
$$
\Delta\bar{\tau}_4 = \frac{1}{\kappa Z_3} \ln\left( \frac{Z_3 + 1}{Z_3 - 1} \right), \tag{41}
$$

where $Z_1 = \sqrt{1 + \frac{4\ln N}{\kappa}}, Z_2 = \sqrt{1 + \frac{16}{\kappa \ln N}}, Z_3 = \sqrt{1 + \frac{4}{\kappa \ln N}}$ and $I(T_1) = \frac{N}{\ln N} \frac{N}{N - I(0) + (N/\ln N)}$. In addition, let

$$
\theta_1^j = \kappa \frac{I_j}{2N} + \frac{1}{2}\sqrt{\left( \frac{\kappa I_j}{N} \right)^2 + \frac{4\kappa}{\ln N}}, \tag{42}
$$
$$
\theta_2^j = \kappa \frac{I_j}{2N} - \frac{1}{2}\sqrt{\left( \frac{\kappa I_j}{N} \right)^2 + \frac{4\kappa}{\ln N}}, \tag{43}
$$

$\Delta\theta_j = \theta_1^j - \theta_2^j$ and

$$
b_j = \frac{N\theta_{1,j} - \kappa I(T_{j-1})}{\kappa I(T_{j-1}) - N\theta_{2,j}}. \tag{44}
$$

Note that, in the above definition, in fact $I(T_{j-1}) = I_{j-1}$ for $j = 3$ and $4$.

Furthermore, for $j = 2, 3$ and $4$, let

$$
d_j = (b_j + \exp(\Delta\theta_j \Delta\bar{\tau}_j)) \tag{45}
$$
$$
q_1^j = \left( \frac{\beta\theta_2^j}{\kappa} - \frac{\beta I_j}{N} \right) \tag{46}
$$
$$
q_2^j = \frac{\beta\theta_1^j}{\kappa} - \frac{\beta I_j}{N} \tag{47}
$$

Finally, we are ready to define the bounding processes used in the proof, $\bar{N}_l(t)$ and $\bar{N}_i(t)$. Let $\bar{N}_i(T_1) = N_i(T_1)$. Furthermore, during Phase $j$, let

$$
\frac{d\bar{N}_i(t)}{dt} = \frac{\rho\bar{N}_l(t) + \kappa\bar{N}_i(t)}{N}(I_j - (\bar{N}_l(t) + \bar{N}_i(t))). \tag{48}
$$

Similarly, let $\bar{N}_l(T_1) = N_l(T_1)$ and, during Phase $j$,

$$
\frac{d\bar{N}_l(t)}{dt} = \begin{cases} C_N + \beta\bar{N}_l(t)\frac{I_j - (\bar{N}_l(t) + \bar{N}_i(t))}{N}, & \bar{N}_w(t) > 0, \\ 0, & \bar{N}_w(t) = 0. \end{cases} \tag{49}
$$

where $\bar{N}_w(t) = I_j - (\bar{N}_i(t) + \bar{N}_l(t))$. Finally, let

$$
\bar{U}(t) = \bar{N}_l(t) + \bar{N}_i(t).
$$

To state the result, we use a bit more notation about these processes. Let $\bar{N}_l^1 = N_l(T_1)$ and for $j = 2, 3,$ and $4$ define

$\bar{N}_l(T_j)$ recursively as follows:

$$
\bar{N}_l^j = \bar{N}_l^{j-1}\left( \frac{1 + b_j}{d_j} \right)^{\frac{\beta}{\kappa}} e^{(-q_1^j \Delta\bar{\tau}_j)} +
$$
$$
+ C_N\left( \frac{b_j}{d_j} \right)^{\frac{\beta}{\kappa}} e^{(-q_1^j \Delta\bar{\tau}_j)} \left( \frac{e^{\left( q_1^j \frac{\ln b_j}{\Delta\theta_j} \right)}}{q_1^j} - \frac{1}{q_1^j} \right) \mathbf{1}_{b \geq 1}
$$
$$
+ C_N\left( \frac{1}{d_j} \right)^{\frac{\beta}{\kappa}} e^{(-q_1^j \Delta\bar{\tau}_j)} \left( \frac{e^{(q_2^j \Delta\bar{\tau}_j)}}{q_2^j} - \frac{e^{\left( q_2^j \frac{\ln b_j}{\Delta\theta_j} \right)}\mathbf{1}_{b \geq 1}}{q_2^j} \right)
$$
$$
- C_N\left( \frac{1}{d_j} \right)^{\frac{\beta}{\kappa}} e^{(-q_1^j \Delta\bar{\tau}_j)} \frac{1}{q_2^j}(1 - \mathbf{1}_{b \geq 1}), \tag{50}
$$

where $\mathbf{1}_{b \geq 1}$ is given by

$$
\mathbf{1}_{b \geq 1} = \begin{cases} 1 & b \geq 1, \\ 0 & b < 1. \end{cases} \tag{51}
$$

We can now state our result characterizing the number of legal and illegal copies at the end of Phases 2-4.

**Lemma 9.** *In the presence of an inefficient, illicit P2P, the number of illegal and legal copies at the end of Phase $j$, $j \in \{2, 3, 4\}$ of the approximate Bass model are given by*

$$
N_l(T_j) \geq \bar{N}_l^j,
$$

*where equality holds when $\beta = 0$.*

From the approximate Bass model (13), the evolution of demand in Phase $j$, for $j = 2, 3$ and $4$, is given by,

$$
I(t) = I_j, \quad \text{where} \quad t \in [T_{j-1}, T_j).
$$

Note that in these three phases, a change in the number of interested copies occurs only at the beginning of the phase and then, it remains constant throughout the phase. That means, the dynamics of evolutions of $N_l(t)$ and $N_i(t)$ in these phases are similar to that of Flash Crowd model discussed in Lemma 5. Also, it can be shown that each of these phases is long enough so that every interested user appearing at the beginning of a phase is being served by the end of that phase. Therefore, we can analyze each of these phases independently. Now, by recursively applying the analysis of Lemma 5 for each of the three phases, we get Lemma 9. A detailed proof of the above lemma is given below.

*Proof:* From the approximate Bass model (13), the evolution of demand in Phase $j$ is,

$$
I(t) = I_j, \quad \text{where} \quad t \in (T_{j-1}, T_j],
$$

and the number of Wanters in Phase $j$ is $N_w(t) = I_j - (N_l(t) + N_i(t))$.

Recall that the efficiency factor of an inefficient illicit P2P, $\eta(t)$, is given by

$$
\eta(t) = \frac{N_r(t) + N_f(t)}{N} = \frac{\rho N_l(t) + \kappa N_i(t)}{N}. \tag{52}
$$

The second equality follows from (5) and (6).

From equation (8), the illegal growth rate in Phase $j$ is

$$\frac{dN_i(t)}{dt} \overset{(a)}{=} \min\{\eta(t)N_w(t), N_r(t) + N_f(t)\},$$

$$\overset{(b)}{=} \eta(t)N_w(t) \tag{53}$$

$$\overset{(c)}{=} \frac{\rho N_l(t) + \kappa N_i(t)}{N}(I_j - (N_l(t) + N_i(t))). \tag{54}$$

Here (a) follows from the fact that $I(t)$ is constant in the last three phases. (b) follows from the definition of $\eta(t)$ and the fact that $N_w(t) \leq N$. (c) follows from (52).

From equation (9), the growth rate of legal copies in Phase $j$ is given by

$$\frac{dN_l(t)}{dt} = \begin{cases} C_N + \beta N_l(t), & N_w(t) > 0, \\ 0, & N_w(t) = 0. \end{cases} \tag{55}$$

The second equality follows from the fact that $\frac{dN_l}{dt} = 0$ when there are no Wanters in the system (from (53)) and $I(t)$ is constant.

Let $U(t)$ be the total copies of the content in the system. Then,

$$U(t) = N_l(t) + N_i(t).$$

Note that the growth rate $N_l(t)$ is at least equal to $C_N$ when $N_w(t) > 0$. In that case, it can be shown that

$$C_N \times (T_j - T_{j-1}) > (I(T_j) - I(T_{j-1}).$$

since $I(0) << C_N$, by assumption. This means that every interested user generated in any one of the last three phases can be served within that phase itself. Furthermore, Lemma 7 shows that no Wanters are left unserved after Phase 1. Therefore, we can conclude that

$$N_l(T_j) + N_i(T_j) = U(T_j) = I(T_j) = I_j. \tag{56}$$

The same arguments hold true in the case of $\bar{N}_l(t)$, i.e,

$$\bar{N}_l(T_j) + \bar{N}_i(T_j) = \bar{U}(T_j) = I(T_j) = I_j. \tag{57}$$

Now, we claim that,

$$N_l(T_j) \geq \bar{N}_l(T_j), \tag{58}$$

and the equality holds when $\beta = 0$.

We can derive $\frac{dN_i}{dU}$ and $\frac{d\bar{N}_i}{dU}$ from the pair of equations (53), (55) and (48), (49) respectively. Then, it can be shown that

$$\frac{dN_i}{dU}\Big|_{N_i=x, U=y} \leq \frac{d\bar{N}_i}{dU}\Big|_{\bar{N}_i=x, U=y}, \tag{59}$$

and the equality holds when $\beta = 0$. Note that the range space of functions $U(t)$ and $\bar{U}(t)$ are identical; in fact they are equal to $[I(T_{j-1}), I(T_j)]$ in Phase $j$ which follows from (56) and (57). Furthermore, recall that the initial values of $N_i(T_1)$ and $\bar{N}_i(T_1)$ are equal by definition. Hence, the conclusion is

$$N_i(T_j) \leq \bar{N}_i(T_j).$$

Then, the claim in (58) is true from the facts that $N_l(T_j) = I(T_j) - N_i(T_j)$ and $\bar{N}_l(T_j) = I(T_j) - \bar{N}_i(T_j)$.

Our objective is to derive an expression of $\bar{N}_l(t)$. Then, evaluate the expression at $t = T_j$ in order to obtain a lower bound on the number of legal copies at the end of each Phase $j$.

Let $\bar{\tau}_j$ be the time such that $\bar{U}(\bar{\tau}_j) = I_j$. This event happens within Phase $j$ itself (from (57)). i.e, $\bar{\tau}_j \in (T_{j-1}, T_j]$. In addition,

$$\bar{N}_w(t) = 0 \text{ when } t \in (\bar{\tau}_j, T_j].$$

Adding (49) and (48), for $t \in (T_{j-1}, \bar{\tau}_j]$, we get,

$$\frac{d\bar{U}}{dt} = ((\beta + \rho)\bar{N}_l(t) + \kappa\bar{N}_i(t))\frac{(I_j - (\bar{N}_l(t) + \bar{N}_i(t)))}{N}$$

$$\overset{(e)}{=} (\kappa\bar{N}_l(t) + \kappa\bar{N}_i(t))\frac{(I_j - (\bar{N}_l(t) + \bar{N}_i(t)))}{N}$$

$$\overset{(f)}{=} \kappa\bar{U}(t)\frac{I_j - \bar{U}(t)}{N}.$$

(e) follows from the fact that $\rho + \beta = \kappa$. (f) follows from the definition of $\bar{U}(t)$ in Phase $j$.

The differential equation given above is a standard Riccatti equation. Its solution is given by

$$\bar{U}(t) = \frac{N\theta_{2,j}}{\kappa} + \frac{N\Delta\theta_j/\kappa}{1 + b_j e^{-\Delta\theta_j(t - T_{j-1})}}, \tag{60}$$

where $\Delta\theta_j = \theta_{1,j} - \theta_{2,j}$. $\theta_{1,j}, \theta_{2,j}$ and $b_j$ are given by equations (42), (43) and (44) respectively.

Let $\Delta\bar{\tau}_j = \bar{\tau}_j - T_{j-1}$. Recall that $\bar{\tau}_j$ is the solution of the equation $\bar{U}(\bar{\tau}_j) = I_j$. Hence, from the above result, we get,

$$\bar{\tau}_j - T_{j-1} = \frac{1}{\Delta\theta_j}\ln\left(\frac{\sqrt{1 + \frac{4}{\kappa\ln N}}j + 1 - \frac{2I(T_{j-1})}{I(T_j)}}{\sqrt{1 + \frac{4}{\kappa\ln N}}j - 1 + \frac{2I(T_{j-1})}{I(T_j)}}\right)$$

$$+ \frac{1}{\Delta\theta_j}\ln\left(\frac{\sqrt{1 + \frac{4}{\kappa\ln N}}j + 1}{\sqrt{1 + \frac{4}{\kappa\ln N}}j - 1}\right). \tag{61}$$

The above expression yields (39), (40) and (41) respectively, when $I(T_j)$ is substituted by actual values from the bass model.

Now, applying the above expression in (49), for $t \in (T_{j-1}, \bar{\tau}_j]$, we get

$$\frac{d\bar{N}_l(t)}{dt} = C_N + \beta\bar{N}_l(t)\frac{I_j - (\bar{N}_l(t) + \bar{N}_i(t))}{N}.$$

A lower bound on the solution of the above differential equation is provided by Lemma 16 in Appendix E. It can be shown that $b\exp(-\Delta\theta_j\Delta\bar{\tau}_j) << 1$. Then $\bar{\tau}_j$ satisfies the condition stipulated by that lemma and a lower bound on the number of legal at the end of Phase $j$ can be obtained by evaluating (147) at $t = \bar{\tau}_j$, which yields $\bar{N}_l^j$ in (50). In case $\beta = 0$, (147) is an exact solution of the above differential equation. ∎

As mentioned in the statement of Lemma 9, the inequality is exact in the case of $\beta = 0$. Additionally, in this case, the form of $N_l(T_4)$ simplifies.

**Corollary 10.** *Let $\beta = 0$. In the presence of an inefficient, illicit P2P, the number of illegal and legal copies at the end of Phase 4 of the approximate Bass model is given by*

$$N_l(T_4) = N_l(T_1) + C_N\sum_{j=2}^{4}\Delta\bar{\tau}_j \tag{62}$$

*where $N_l(T_1)$ is given by Corollary 8.*

Now that we have characterized the number of legal and illegal copies at the end of Phase $4$ precisely, attaining the statement in Theorem 1 is accomplished by taking studying the asymptotics of the results in Lemma 9 and Corollary 10. Throughout, we use $A_N \sim B_N$ to denote $\lim_{N \to \infty} \frac{A_N}{B_N} = 1$.

To begin, recall from (10) that,

$$L = \frac{N_l(T_\infty)}{N} = \frac{N_l(T_\infty)}{N} \tag{63}$$

$$\geq \frac{\bar{N}_l^4}{N}, \tag{64}$$

where $\bar{N}_l^4$ is recursively defined by (50) in terms of $\bar{N}_l^1$, $\bar{N}_l^2$ and $\bar{N}_l^3$. As $N$ goes larger, from the above equation, we get that

$$L \in \Omega \left( \frac{\ln \ln N + (\ln N)^{\frac{\beta}{\kappa}}}{\ln N} \right) \tag{65}$$

and $L \in \Theta \left( \frac{\ln \ln N}{\ln N} \right)$ if $\beta = 0$, which completes the proof of Theorem 1.

## Appendix C
## Proof of Theorem 3

The proof of Theorem 3 parallels to that of Theorem 1. We mimic the approach of the proof of Theorem 3 and define two processes $\bar{N}_l(t)$ and $\bar{N}_i(t)$ that bound $N_l(t)$ and $N_i(t)$ and analyze these processes. Importantly, the bounding processes are equivalent to the original processes when $\beta = 0$.

Let $\bar{U}(t) = \bar{N}_l(t) + \bar{N}_i(t)$. Further, let $\bar{N}_l(0) = N_l(0) = 0$ and

$$\frac{d\bar{N}_l(t)}{dt} = = \begin{cases} C_N + \beta\bar{N}_l(t) & \bar{N}_w(t) > 0, \\ 0 & \bar{N}_w(t) = 0. \end{cases} \tag{66}$$

where $\bar{N}_w(t) = N - \bar{U}(t)$. Furthermore, we define $\bar{N}_i(0) = N_i(0) = 0$ and

$$\frac{d\bar{N}_i(t)}{dt} = \begin{cases} \rho\bar{N}_l(t) + \kappa\bar{N}_i(t) & 0 \leq \bar{U}(t) \leq \frac{N}{1+\rho}, \\ N - \bar{N}_l(t) - \bar{N}_i(t) & \frac{N}{1+\rho} \leq \bar{U}(t) \leq N. \end{cases} \tag{67}$$

Finally, let $\bar{N}_i(0) = N_i(0) = 0$. To state the results, we may need a bit more notation. Let

$$\bar{N}_l = \frac{N}{\ln N\beta} \left( e^{\beta\bar{\tau}} - 1 \right). \tag{68}$$

Furthermore, $\bar{\tau} = \frac{1}{1+\beta} \ln \left( 1 + \frac{\ln N(1+\beta)H^{\frac{-\beta}{\kappa}}}{1+\rho} \right) + \frac{1}{\kappa} \ln(H)$, where $H = 1 + \frac{\kappa \ln N}{(1+\rho)}$. Now, we characterize the number of legal copies and illegal copies in the following lemma.

**Lemma 11.** *In the presence of an efficient, illicit P2P, the number of illegal copies is given by*

$$N_l(T_\infty) \geq \bar{N}_l, \tag{69}$$

*and the equality holds when $\beta = 0$.*

*Proof:*

From equation (8), the growth rate of illegal copies is given by

$$\frac{dN_i}{dt} \overset{a}{=} \min\{N_w(t), \rho N_l(t) + \kappa N_i(t)\} \tag{70}$$

$$\overset{b}{=} \min\{I(t) - U(t), \rho N_l(t) + \kappa N_i(t)\} \tag{71}$$

where (a) follows from equations (5), (6) along with the facts that $\eta = 1$ and $I(t)$ is constant. (b) follows from the definition of the number of wanters in the system.

From equation (9), the growth rate of legal copies in Phase $j$ is given by

$$\frac{dN_l(t)}{dt} \overset{c}{=} C_N + \beta N_l(t) \quad \text{if} \quad N_w(t) > 0,$$

$$\overset{d}{=} 0 \quad \text{if} \quad N_w(t) = 0. \tag{72}$$

(d) follows from the facts that $\frac{dN_l}{dt} = 0$ when there are no wanters in the system (from (70)) and $I(t)$ is constant.

As defined before, let $U(t)$ be the total copies of the content in the system. Then, $U(t) = N_l(t) + N_i(t)$.

Now, we claim that,

$$N_l(T_j) \geq \bar{N}_l(T_j). \tag{73}$$

and the equality holds when $\beta = 0$.

Note that

$$\frac{d\bar{N}_l(t)}{dt}\Big|_{\bar{U}=x, \bar{N}_i=y} \overset{e}{=} \frac{dN_l(t)}{dt}\Big|_{U=x, N_i=y}, \tag{74}$$

$$\frac{d\bar{N}_i(t)}{dt}\Big|_{\bar{U}=x, \bar{N}_i=y} \overset{f}{\geq} \frac{dN_i(t)}{dt}\Big|_{U=x, N_i=y}. \tag{75}$$

and (f) is an equality when $\beta = 0$. (e) follows from (66) and (72). And (f) is due to (70) and (67). From the above equations, we can deduce that

$$\frac{d\bar{N}_l}{d\bar{U}}\Big|_{\bar{U}=x, \bar{N}_i=y} \leq \frac{dN_l}{dU}\Big|_{U=x, N_i=y}. \tag{76}$$

Note that the range of functions $U(t)$ and $\bar{U}(t)$ are identical, $[I(0), N]$. Since $N_l(0) = \bar{N}_l(0)$, from the above equation, we get that $N_l(T_j) \geq \bar{N}_l(T_j)$, Also, equality holds when $\beta = 0$.

Let $\bar{\tau}$ be the instant at which $\bar{N}_w(\bar{\tau}) = 0$. Then, the number of legal copies, $N_l(t)$, is given by

$$\bar{N}_l(t) = \begin{cases} \left( \frac{C_N}{\beta} \right) e^{\beta t} - \frac{C_N}{\beta} & t \in (0, \bar{\tau}], \\ \bar{N}_l(\bar{\tau}) & t > \bar{\tau}. \end{cases} \tag{77}$$

The above result follows from (66) and the initial condition $N_l(0) = 0$. Now, we resort to find $\bar{\tau}$. Note that, $\bar{N}_w(\bar{\tau}) = 0$ implies $\bar{U}(\bar{\tau}) = N$. Therefore, first we derive $\bar{U}(t)$ and then, finds the time at which $\bar{U}(t)$ reaches $N$.

Note that $\bar{U}(0) < \frac{N}{1+\rho}$, by assumption. Then, from (66) and (67), we get that

$$\frac{d\bar{U}(t)}{dt} = \rho\bar{U}(t) + C_N, \quad \text{if} \quad t \in [0, \nu],$$

where $\nu$ is defined as $\bar{U}(\nu) = \frac{N}{1+\rho}$. Solving the above equation with the initial condition $\bar{U}(0) = 0$ yields

$$\bar{U}(t) = \frac{C_N}{\kappa} e^{\kappa t} - \frac{C_N}{\kappa}, \quad \text{if} \quad t \in [0, \nu]. \tag{78}$$

Then, from the above result $\nu$ can shown to be $\nu = \frac{1}{\kappa} \ln(H)$, where $H = 1 + \frac{\kappa \ln N}{1+\rho}$.

Now, consider the case $t \in [\nu, \bar{\tau}]$. Then, $\frac{N}{1+\rho} \leq \bar{U}(t) \leq N$ and hence, from (67),

$$\frac{dN_i}{dt} = N - \bar{N}_l(t) - \bar{N}_i(t), \quad \text{if} \quad t \in [\nu, \bar{\tau}].$$

Solving the above equation, we get

$$
\begin{aligned}
\bar{N}_i(t) &= N - \left(\bar{N}_l(\nu) + \frac{C_N}{\beta}\right)\frac{e^{\beta(t-\nu)}}{1+\beta} + \frac{C_N}{\beta} \\
&\quad + \left(\bar{N}_i(\nu) + \frac{\bar{N}_l(\nu)}{1+\beta} - \frac{C_N}{1+\beta} - N\right)e^{-(t-\nu)}, \\
&= N - \frac{C_N}{\beta}\frac{e^{\beta(t)}}{1+\beta} + \frac{C_N}{\beta} \\
&\quad - \left(\frac{N\rho}{1+\rho} + \frac{C_N e^{\beta\nu}}{1+\beta}\right)e^{-(t-\nu)},
\end{aligned}
$$

for $t \in [\nu, \bar{\tau}]$. Here, the second equality is obtained by replacing $\bar{N}_i(\nu)$ with $\bar{U}(\nu) - \bar{N}_l(\nu)$ and by substituting $\bar{N}_l(\nu)$ from (77). Then, $\bar{U}(t)$, which is eaqul to $\bar{N}_l(t) + \bar{N}_i(t)$, is given by

$$
\bar{U}(t) = N + \frac{C_N e^{\beta t}}{1+\beta} - \left(\frac{N\rho}{1+\rho} + \frac{C_N e^{\beta\nu}}{1+\beta}\right)e^{-(t-\nu)}.
$$

Now, solving for $t$, from $\bar{U}(t) = N$, we get that

$$
\bar{\tau} = \nu + \frac{1}{1+\beta}\ln\left(1 + \frac{\ln N(1+\beta)e^{-\beta\nu}}{1+\rho}\right) \tag{79}
$$

$$
= \frac{1}{\kappa}\ln H + \frac{1}{1+\beta}\ln\left(1 + \frac{\ln N(1+\beta)H^{\frac{-\beta}{\kappa}}}{1+\rho}\right). \tag{80}
$$

The second result follows by susbtituting $\nu = \frac{1}{\kappa}\ln H$, where $H = 1 + \frac{\kappa \ln N}{1+\rho}$.

Finally, substituting $\bar{\tau}$ in (77) yields $\bar{N}_l$, which completes the proof. ∎

As mentioned in the statement of Lemma 11, the inequality is exact in the case of $\beta = 0$. Additionally, in this case, the form of $N_l(T_\infty)$ simplifies.

**Corollary 12.** *Let $\beta = 0$. Then, the number of legal copies at the end of Phase 4 is given by $N_l(T_\infty) = C_N \bar{\tau}$.*

Now that we have characterized the number of legal and illegal copies precisely, attaining the statement in Theorem 3 is accomplished by studying the asymptotics of the results in Lemma 11 and Corollary 12. From (10), Lemma 11, Corollary 12 and equation (68), we can show that

$$
L \in \Omega\left(\frac{1}{\ln N}\frac{(\ln N)^{\frac{\beta}{\kappa}} - 1}{\left(\frac{\beta}{\kappa}\right)}\right), \tag{81}
$$

and $L \in \Theta\left(\frac{\ln\ln N}{\ln N}\right)$ if $\beta = 0$, which completes the proof.

## APPENDIX D
## PROOF OF THEOREM 4

In our model, an efficient illicit P2P is characterized by efficiency parameter, $\eta(t)$, equal to one. Then, from (8), the evolution of illegal copies of content in the system, $N_i(t)$, is given by

$$
\frac{dN_i(t)}{dt} = \min\left\{N_w(t) + \frac{dI(t)}{dt}, \rho N_l(t) + \kappa N_i(t)\right\}. \tag{82}
$$

And, the evolution of legal copies of the content in the system, $N_l(t)$, is given by,

$$
\frac{dN_l(t)}{dt} = \begin{cases} C_N + \beta N_l(t) & N_w(t) > 0, \\ \min\{C_N + \beta N_l(t), \frac{dI}{dt} - \frac{dN_i}{dt}\} & N_w(t) = 0. \end{cases} \tag{83}
$$

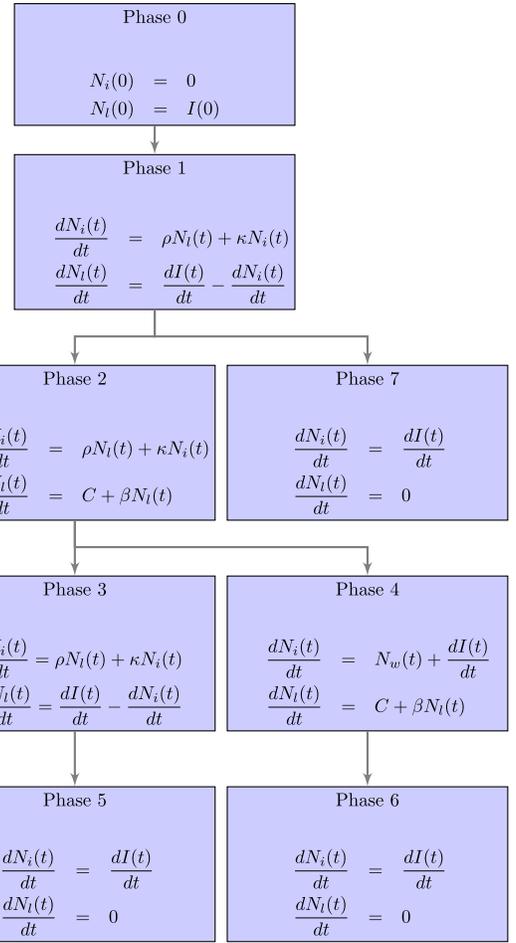

Fig. 6. Evolutionary phases of the growth of Legal and Illegal copies of content in the presence of an efficient Illicit P2P

As the interest for the content evolves according to the Bass demand model, the evolution of $N_l(t)$ and $N_i(t)$ traverses along multiple stages of dynamics as shown in Figure 6. Below, we discuss these stages of evolution in detail.

**Stage 1**: By assumption, $N_l(0) = I(0)$, $N_i(0) = 0$ and $N_w(0) = 0$ where $I(0)$ is the initial demand in the system. Then,

$$
N_w(0) + \frac{dI(t)}{dt}|_{t=0} > \rho N_l(0) + \kappa N_i(0).
$$

The above result follows from our assumption that $\kappa < 1 - \frac{I(0)}{N}$. Therefore, at $t = 0$, from (82),

$$
\frac{dN_i(t)}{dt} = \rho N_l(t) + \kappa N_i(t). \tag{84}
$$

From (83), the evolution of $N_l(t)$ at time $t = 0$ is,

$$
\begin{aligned}
\frac{dN_l(t)}{dt} &= \frac{dI(t)}{dt} - \frac{dN_i(t)}{dt}, \\
&= \frac{dI(t)}{dt} - (\rho N_l(t) + \kappa N_i(t)). \tag{85}
\end{aligned}
$$

The first equality follows from the facts that $N_w(0) = 0$ and $\frac{dI(t)}{dt}|_{t=0} < C_N$. Also, from the above equations, we get that $N_l(t) + N_i(t) = I(t)$.

The evolution exits Stage 1 when any one of the following conditions is attained,

$$\text{C1}: \qquad \frac{dI}{dt}(t) - \frac{dN_i}{dt} \geq C_N + \beta N_l(t), \qquad (87)$$

$$\text{C2}: \qquad \frac{dI}{dt}(t) \leq \rho N_l(t) + \kappa N_i(t). \qquad (88)$$

Here, C1 occurs when the number of wanters approaching the legitimate CDN exceeds its current capacity, Then, from (83), the dynamics of evolution of $N_l(t)$ changes. C2 happens when the number of users attempting to download from the illicit P2P reduces the current capacity of the illicit P2P. Then, from (82), the dynamics of evolution of $N_i(t)$ changes. Next, we show if $\kappa < 1 - \frac{2}{\sqrt{\ln N}}$, C1 occurs before C2 and the evolution proceeds to Stage 2. Otherwise, Stage 1 is followed by Stage 7.

Now, let $T_2$, be the time at which C1 is attained, i.e,

$$\frac{dI}{dt}\mid_{t=T_2} - \frac{dN_i}{dt}\mid_{t=T_2} = C_N + \beta N_l(T_2), (89)$$

$$\Rightarrow \quad \frac{dI}{dt}\mid_{t=T_2} - \kappa I(T_2) = C_N \qquad (90)$$

$$\Rightarrow \quad I(T_2) = \frac{N(1-\kappa)}{2}\left[1 - \sqrt{1 - \frac{4}{\ln N(1-\kappa)^2}}\right] (91)$$

The second equality follows from (84) along with the facts that $\kappa = \rho + \beta$ and $N_l(t) + N_i(t) = I(t)$. Equation (91) follows from the definition of $I(t)$. In the above equation, $T_2$ has a real positive solution iff $\kappa < 1 - \frac{2}{\sqrt{\ln N}}$. Also, let $T_7$ be the time at which C2 is attained, i.e,

$$\frac{dI}{dt}\mid_{t=T_7} = \rho N_l(T_7) + \kappa N_i(T_7)$$
$$\Rightarrow \frac{dI}{dt}\mid_{t=T_7} - \kappa I(T_7) = -\beta N_l(T_7). \qquad (92)$$

The second equality follows from the facts that $\kappa = \rho + \beta$ and $N_l(t) + N_i(t) = I(t)$. From (90), (92) and the definition of $I(t)$, it can be shown that, if $T_2$ has a real valued solution, then $T_2 < T_7$. Therefore, Stage 1 is followed by Stage 2 if $\kappa < 1 - \frac{2}{\sqrt{\ln N}}$ and, Stage 7 otherwise.

**Stage 2**: The evolution enters Stage 2 from Stage 1 due to the condition C1 given by (87). Then, the dynamics of $N_i(t)$ does not change from that of Stage 1,

$$\frac{dN_i}{dt} = \rho N_l(t) + \kappa N_i(t), \qquad (93)$$

but the dynamics of $N_l(t)$ changes to,

$$\frac{dN_l}{dt} = C_N + \beta N_l(t). \qquad (94)$$

Also, from the above equations and (87), $N_l(t) + N_i(t) \leq I(t)$.

A transition from this stage occurs when any one of the following conditions is satisfied,

$$\text{C3}: \qquad C_N + \beta N_l(t) \geq \frac{dI(t)}{dt} - \frac{dN_i(t)}{dt},$$
$$N_w(t) = 0, \qquad (95)$$

$$\text{C4}: \qquad \frac{dI(t)}{dt} + N_w(t) \leq \rho N_l(t) + \kappa N_i(t). \qquad (96)$$

Here, C3 occurs when the number of wanters in the system goes to zero and the rate at which newly generated population approaching the legitimate CDN falls below its current capacity. Then, from (83), the dynamics of evolution of $N_l(t)$ changes. C2 happens when the number of users attempting to download from the illicit P2P reduces below the current capacity of the illicit P2P. Then, from (82), the dynamics of evolution of $N_i(t)$ changes. The evolution enters Stage 3, if C3 is attained before C4. Otherwise, it proceeds to Stage 4.

Let $T_3$ mark the time at which the evolution enters Stage 3. Then, from C3 and (93),

$$C_N + \beta N_l(T_3) \geq \frac{dI(t)}{dt}\mid_{t=T_3} - (\rho N_i(T_3) + \kappa N_l(T_3)), \quad (97)$$

$$\text{and} \quad N_w(T_3) = 0. \qquad (98)$$

Also, let Stage 4 start at time $t = T_4$. Then, from C4,

$$\frac{dI(t)}{dt}\mid_{t=T_4} + N_w(T_4) = \rho N_l(T_4) + \kappa N_i(T_4). \qquad (99)$$

**Stage 3**: The evolution enters Stage 3 from Stage 2 due to the condition C3 given by (95). Then, the dynamics $N_i(t)$ does not change from that of Stage 2,

$$\frac{dN_i(t)}{dt} = \rho N_l(t) + \kappa N_i(t), \qquad (100)$$

but, the evolution of $N_l(t)$ changes to,

$$\frac{dN_l(t)}{dt} = \frac{dI(t)}{dt} - \frac{dN_i(t)}{dt}, \qquad (101)$$
$$= \frac{dI(t)}{dt} - (\rho N_l(t) + \kappa N_i(t)). \qquad (102)$$

This stage starts at $t = T_3$, which is defined by (97) and (98). From the above dynamics equations and (98), we get $N_l(t) + N_i(t) = I(t)$.

We show that the evolution of $N_l(t)$, given by (101), does not change as long as the evolution of $N_i(t)$ does not deviate from (100). This claim holds true if

$$C_N + \beta N_l(t) \geq \frac{dI(t)}{dt} - (\rho N_l(t) + \kappa N_i(t)),$$
$$\Rightarrow \frac{dI(t)}{dt} - \kappa I(t) \leq C_N, \qquad (103)$$

for all $t \geq T_3$. The second inequality follows from the facts $\kappa = \rho + \beta$ and $N_l(t) + N_i(t) = I(t)$. At $t = T_3$ the above requirement is met, which follows from (97). Then, we get

$$I(T_3) \geq \frac{N(1-\kappa)}{2}, \qquad (104)$$

from the definition of $I(t)$ and (103). The function $\frac{dI(t)}{dt} - \kappa I(t)$ is monotonically decreasing if $I(t) > \frac{N(1-\kappa)}{2}$. Then, (103) holds for all $t > T_3$ and that proves our claim.

The above discussion implies that a transition from this stage happens only when the dynamics of evolution of $N_i(t)$ changes. From (82) and (100), the dynamics of $N_i(t)$ changes, when the number of users downloading from the illicit P2P reduces below the current capacity of illicit P2P,

$$\text{C5}: \quad \frac{dI(t)}{dt} \leq \rho N_l(t) + \kappa N_i(t). \qquad (105)$$

When C5 occurs, evolution enters Stage 5. Let this occurs at $t = T_5$. Then,

$$\frac{dI(t)}{dt}\mid_{t=T_5} = \rho N_l(T_5) + \kappa N_i(T_5). \qquad (106)$$

**Stage 4**: The evolution enters Stage 3 from Stage 2 due to the condition C4 given by (96). Then, the dynamics of $N_l(t)$ does not change from that of Stage 2,

$$\frac{dN_l(t)}{dt} = C_N + \beta N_l(t), \tag{107}$$

but the evolution of $N_i(t)$ changes to,

$$\frac{dN_i(t)}{dt} = N_w(t) + \frac{dI(t)}{dt}, \tag{108}$$

This stage starts at time $t = T_4$ defined by (99).

We claim that the evolution of $N_i(t)$ follows (108) for all $t \geq T_4$. This claim holds true if

$$\left(N_w(t) + \frac{dI(t)}{dt}\right) \leq \rho N_l(t) + \kappa N_i(t), \tag{109}$$

for all $t \geq T_4$. Note that Equation (109) holds true at $t = T_4$. Since, $N_w(t) = I(t) - (N_l(t) + N_i(t))$ by definition, from Equation (108), we get that $\frac{dN_w(t)}{dt} < 0$. Also, using the definition of $N_w(t)$ in (99), we can show that

$$\frac{dI(t)}{dt}\big|_{t=T_4} - \kappa I(T_4) = -(1+\kappa)N_w(T_4) - \beta N_l(T_4) < 0.$$

Then, from the definition of $I(t)$, the above result holds for all $t \geq T_4$. Then, we get

$$\frac{d}{dt}\left(N_w(t) + \frac{dI}{dt}\right) < \frac{d}{dt}(\rho N_l(t) + \kappa N_i(t)),$$

which along with (99) proves (109).

The above discussion implies that a transition from this stage occurs when the evolution of $N_l(t)$ changes. From (107) and (83), the evolution of $N_l(t)$ changes when the number of wanters goes to zero. Then,

$$N_w(T_6) = 0. \tag{110}$$

where $T_6$ marks the beginning of Stage 6.

**Stage 5, 6, 7**:
These are the final stages of evolution. Stage 5 is preceded by Stage 3, Stage 6 is preceded by Stage 4, and Stage 7 is preceded by Stage 1. The dynamics of all these stages are identical,

$$\frac{dN_l(t)}{dt} = 0, \tag{111}$$

$$\frac{dN_i(t)}{dt} = \frac{dI(t)}{dt}. \tag{112}$$

It is easy to see that the evolutions of $N_l(t)$ and $N_l(t)$ stay in these stages forever once they reach here.

In summary, if $\kappa \geq 1 - \frac{2}{\sqrt{\ln N}}$, the evolution of $N_i(t)$ and $N_l(t)$ traverse the sequence of phases, *Stage* 1 →*Stage* 7. Otherwise, they proceed along the sequence of phases, *Stage* 1 → *Stage* 2 →*Stage* 3(*Stage* 4) →*Stage* 5(*Stage* 6). In the next section, we analyze these two cases separately and obtain a lower bound on number of legal copies of the content in the system at the end of evolution.

### A. Analysis

We first consider the case, $\kappa \geq 1 - \frac{2}{\sqrt{\ln N}}$. Let us introduce a few notation before stating the result. We define

$$\Phi(x) = \left(\frac{I(0)}{N}\right)^{\beta} N \left[(1-\kappa)\psi\left(\beta, \frac{x}{N}\right) - \kappa\psi\left(\beta - 1, \frac{x}{N}\right)\right], \tag{113}$$

and $\psi(\beta, x) = \int_{I(0)/N}^{x} \left(\frac{1-u}{u}\right)^{\beta} du$. Also, let

$$\bar{T} = \ln\left[\frac{N(1-\kappa)G}{I(0)(2 - (1-\kappa)G)}\right], \tag{114}$$

where $G = 1 + \sqrt{1 + \frac{4\beta D}{N(1-\kappa)^2}}$ and $D = \Phi(N(1-\kappa))\left(\frac{N(1-\kappa)}{I(0)\kappa}\right)^{\beta}$. Now, we are ready to provide the result.

**Lemma 13.** *Assume* $\kappa \geq 1 - \frac{2}{\sqrt{\ln N}}$. *Then, a lower bound on the number of legal copies of the content in the system at* $t = T_\infty$ *is given by,*

$$N_l(T_\infty) \geq (\Phi(I(\bar{T})) + I(0))e^{\beta \bar{T}}. \tag{115}$$

*where* $I(t)$ *is given by (3).*

*Proof:* Recall that, when $\kappa \geq 1 - \frac{2}{\sqrt{\ln N}}$, the evolution of $N_l(t)$ and $N_i(t)$ takes place in two stages, namely Stage 1 and Stage 7. Solving the dynamics of evolution in Stage 1, given by (85) and (84), we get

$$\begin{aligned} N_l(t) &= (\Phi(I(t)) - \Phi(I(0)))e^{\beta t} + I(0)e^{\beta t}, \\ &= (\Phi(I(t)) + I(0))e^{\beta t}, \end{aligned} \tag{116}$$

where $\Phi(x)$ is defined by (113). The second equality follows since $\Phi(I(0)) = 0$.

Stage 7 starts at $t = T_7$. Recall from (92) that $T_7$ is a solution to the equation,

$$\frac{dI(t)}{dt} - \kappa I(t) = -\beta N_l(t)$$

. It is not easy to solve the above equation exactly . Hence, here, we obtain a lower bound on $T_7$. Let $r = \ln(\frac{N(1-\kappa)}{I(0)\kappa})$. Note that, at $t = r$,

$$\frac{dI}{dt}(t) - \kappa I(t) = 0.$$

Also, the function $\frac{dI}{dt}(t) - \kappa I(t)$ is positive for $t < r$ and, it is monotonically decreasing for $t \geq r$. Then, $r \leq T_7$. Then, $N_l(r) \leq N_l(T_7)$. That implies the solution of the equation,

$$\frac{dI}{dt} - \kappa I = -\beta N_l(r),$$

must be less than or equal to $T_7$. Now, substituting $N_l(r)$ from Equation (116) in the above equation, and then, solving for $t$ yields $\bar{T}$, which is defined by (114), as the unique solution. Since no legals are generated in Stage 7 according to (111), and $T_7 \geq \bar{T}$, we have

$$N_l(T_\infty) = N_l(T_7) \geq N_l(\bar{T}).$$

Now, obtain $N_l(\bar{T})$ from (116) and substitute in the above inequality to prove the lemma. ∎

Now, we consider the second case where $\kappa < 1 - \frac{2}{\sqrt{\ln N}}$. We introduce a few notation before stating the result. Let

$$I_2 = \frac{N(1-\kappa)}{2}\left[1 - \sqrt{1 - \frac{4}{\ln N(1-\kappa)^2}}\right], \quad (117)$$

$$T_2 = \ln\left[\frac{NI_2}{I(0)(N-I_2)}\right], \quad (118)$$

$$I_3 = \frac{I_2 e^{\Delta T_1}}{1 - \frac{I_2}{N} + \frac{I_2}{N}e^{\Delta T_1}}, \quad (119)$$

$$\Delta T_1 = \frac{1}{\kappa}\ln\left[\frac{\frac{c}{\kappa} + \frac{N(1-\kappa)}{2}[1+H]}{\frac{c}{\kappa} + \frac{N(1-\kappa)}{2}[1-H]}\right], \quad (120)$$

$$\Delta T_2 = \frac{1}{\kappa}\ln\left[\frac{\frac{c}{\kappa} + I_3}{\frac{c}{\kappa} + I_2}\right], \quad (121)$$

$$\bar{T}_3 = T_2 + \Delta T_2 \quad (122)$$

$$L_3 = \frac{C}{\beta}(e^{\beta\Delta T_2} - 1) + (\Phi(I_2) + I(0))e^{\beta\bar{T}_3},$$

where $H = \sqrt{1 - \frac{4}{\ln N(1-\kappa)^2}}$.

Also, let

$$I_4 = I(\bar{T}_3) = \frac{I(0)e^{\bar{T}_3}}{1 - \frac{I(0)}{N} + \frac{I(0)}{N}e^{\bar{T}_3}}, \quad (123)$$

$$I_5 = \frac{N(1-\kappa)}{2}\left[1 + \sqrt{1 + \frac{4\beta L_3}{N(1-\kappa)^2}}\right], \quad (124)$$

$$\bar{T}_5 = \ln\left[\frac{NI_5}{I(0)(N-I_5)}\right], \quad (125)$$

$$L_4 = (\Phi(I_5) - \Phi(I_4))e^{\beta\bar{T}_5} + L_3 e^{\beta(\bar{T}_5 - \bar{T}_3)},$$

where $I(t)$ is the Bass demand function.

**Lemma 14.** *Assume* $\kappa < 1 - \frac{2}{\sqrt{\ln N}}$. *Then, a lower bound on the number of legals at* $t = T_\infty$ *is given by,*

$$N_l(T_\infty) \geq \begin{cases} L_3 & \text{if} \quad \bar{T}_5 \leq \bar{T}_3 \\ L_4, & \text{else.} \end{cases} \quad (126)$$

*Proof:* When $\kappa < 1 - \frac{2}{\sqrt{\ln N}}$, the evolution of of $N_l(t)$ and $N_i(t)$ takes place along a sequence of stages, which is given by, 'Stage 1 → Stage 2 →Stage 3(or Stage 4)→Stage 5(or Stage 6)'. An exact characterization of $N_l(t)$ and $N_i(t)$ might be quite difficult as the analysis involves solving many complex differential equations. Therefore, we define two processes $\bar{N}_l(t)$ and $\bar{N}_i(t)$; $\bar{N}_l(t)$ bounds $N_l(t)$ from below and $\bar{N}_i(t)$ bounds $N_i(t)$ from above. We analyze these bounding processes instead of the actual processes.

We go through a sequence of intermediate steps to prove this lemma.

**Step 1**: *Define* $\bar{N}_l(t)$ *and* $\bar{N}_i(t)$

First of all, let $\bar{N}_l(0) = N_l(0)$ and $\bar{N}_i(0) = N_i(0)$. Let $\bar{N}_l(t)$ evolves as follows,

$$\frac{d\bar{N}_l(t)}{dt} = \begin{cases} \frac{dI}{dt} - (\rho\bar{N}_l(t) + \kappa\bar{N}_i(t)), & [0, T_2], \\ C_N + \beta\bar{N}_l(t), & [T_2, \bar{T}_3], \\ \frac{dI}{dt} - (\rho\bar{N}_l(t) + \kappa\bar{N}_i(t)), & [\bar{T}_3, \max\{\bar{T}_3, \bar{T}_5\}], \\ 0, & [\max\{\bar{T}_3, \bar{T}_5\}, T_\infty]. \end{cases} \quad (127)$$

Also, let

$$\frac{d\bar{N}_i(t)}{dt} = \begin{cases} (\rho\bar{N}_l(t) + \kappa\bar{N}_i(t)), & [0, T_2], \\ (\rho\bar{N}_l(t) + \kappa\bar{N}_i(t)) \\ \quad + R\delta(t - \bar{T}_3), & [T_2, \bar{T}_3], \\ (\rho\bar{N}_l(t) + \kappa\bar{N}_i(t)), & (\bar{T}_3, \max\{\bar{T}_3, \bar{T}_5\}], \\ \frac{dI}{dt} & [\max\{\bar{T}_3, \bar{T}_5\}, T_\infty]. \end{cases} \quad (128)$$

where $T_2$ is given by (118), $\bar{T}_3$ is defined by (122), $\bar{T}_5$ is defined by (125), $R = I(\bar{T}_3) - (N_l(\bar{T}_3) + N_i(\bar{T}_3))$ and $\delta(t)$ is Kronecker delta function. It can be verified that $\bar{T}_3 > T_2$. Also, the following equations can be verified:

$$\frac{dI(t)}{dt}\Big|_{t=\bar{T}_3} - \kappa I(\bar{T}_3) \leq C_N, \quad (129)$$

$$\bar{N}_l(t) + \bar{N}_i(t) < I(t) \quad T_2 < t < \bar{T}_3, \quad (130)$$

$$\frac{dI(t)}{dt}\Big|_{t=\bar{T}_5} - \kappa I(\bar{T}_5) = \beta\bar{N}_l(\bar{T}_3). \quad (131)$$

Also, we define $\bar{N}_w(t) = I(t) - (\bar{N}_l(t) + \bar{N}_i(t))$. In the next step, we show that $\bar{N}_l(t) \leq N_l(t)$ for all $t$.

**Step 2**: *We claim that* $\bar{N}_l(t) \leq N_l(t)$:

Recall that, the actual processes may pass through either Stages 3 and 5 or Stages 4 and 6. We analyze these two cases separately.

*Case* 1: *The evolution of* $N_l(t)$ *and* $N_i(t)$ *takes place along Stages* 3 *and* 5

First of all, we have $N_l(0) = \bar{N}_l(0)$ and $N_i(0) = \bar{N}_i(0)$ from the definition of the bounding processes. Now, suppose $\bar{T}_3 \leq T_3$. Then, comparing Stage 1 dynamics, (85, 84), and Stage 2 dynamics (94, 93) with the bounding process dynamics (127, 128), we get that, for $t \in [0, \bar{T}_3]$,

$$\frac{d\bar{N}_l(t)}{dt} = \frac{dN_l(t)}{dt} \quad \text{and} \quad \frac{d\bar{N}_i(t)}{dt} \geq \frac{dN_i(t)}{dt}.$$

Then,

$$\bar{N}_l(t) = N_l(t) \quad \text{if} \quad t \in [0, \bar{T}_3]. \quad (132)$$

Also, suppose $\bar{T}_5 \leq T_5$. Then, comparing Stage 2 dynamics, (94, 93), Stage 3 dynamics (101, 100) and Stage 5 dynamics (111, 112) with the bounding process dynamics (127, 128), we get that, for $t \in [\bar{T}_3, T_\infty]$,

$$\frac{d\bar{N}_l(t)}{dt} \leq \frac{dN_l(t)}{dt} \quad \text{and} \quad \frac{d\bar{N}_i(t)}{dt} \geq \frac{dN_i(t)}{dt}.$$

Then, $\bar{N}_l(t) \leq N_l(t)$ for $t > \bar{T}_3$. To complete the proof, we must show that $\bar{T}_3 \leq T_3$ and $\bar{T}_5 \leq T_5$.

*Show that* $\bar{T}_3 \leq T_3$: Recall that Stage 3 begins at $T_3$ in the evolution of the original processes. From the definition of $T_3$, given by (97),

$$\frac{dI(t)}{dt}\Big|_{t=T_3} - (\rho N_l(T_3) + \kappa N_i(T_3)) \leq C_N + \beta N_l(T_3),$$

$$\Rightarrow \frac{dI(t)}{dt}\Big|_{t=T_3} - \kappa I(T_3) \leq C_N. \quad (133)$$

The second inequality follows from the facts that $\kappa = \rho + \beta$ and $N_i(T_3) + N_l(T_3) = I(T_3)$ (since $N_w(T_3) = 0$ from (98)).

First, we guess a lower bound for $T_3$. Suppose, at time $t = r$,

$$I(r) = \frac{N(1-\kappa)}{2}\left[1 + \sqrt{1 + \frac{4}{\ln N(1-\kappa)^2}}\right],$$

is satisfied. Note that $I(r) > I(T_2)$ and hence, $r > T_2$. It can be shown that if $t \in [T_2, r]$,

$$\frac{dI}{dt}(t) - \kappa I(t) \geq C_N,$$

with equality at $t = T_2$ and $t = r$. Also, the function, $\frac{dI}{dt}(t) - \kappa I(t)$ strictly decreasing if $t \geq r$. Then, from (133) and the fact that $T_3 > T_2$, we conclude that $r \leq T_3$.

Now, obtain a better lower bound for $T_3$. Let us define $U(t) = N_l(t) + N_i(t)$. From (98), we have $N_w(T_3) = 0$, which implies that $U(T_3) = I(T_3)$. We know that $U(r) \leq I(r)$. Find $t'$ such that $U(t') = I(r)$. Then, $U(t') \leq I(t')$. Then, get $s$ such that $U(s) = I(t')$. Since $U(t)$ and $I(t)$ are monotonically increasing, we have $r \leq t' \leq s \leq T_3$.

From the dynamics of evolution of Stage 2, given by (93) and (94), we can show that during the interval $[T_2, T_3]$,

$$U(t) = \left( \frac{C}{\kappa} + I_2 \right) e^{\kappa(t - T_2)} - \frac{C}{\kappa}.$$

Then, it can be shown that $t' = T_2 + \Delta T_1, I_3 = I(t')$ and $s = \bar{T}_3$. Hence, $\bar{T}_3 \leq T_3$.

*Show that $\bar{T}_5 \leq T_5$:* Recall that Stage 5 begins at $T_5$. From (106),

$$\frac{dI(t)}{dt} \Big|_{t = T_5} - \kappa I(T_5) = -\beta N_l(T_5).$$

The above result is due to the facts that $\kappa = \rho + \beta$ and $N_i(t) + N_l(t) = I(t)$ in Stage 3 and 5.

We guess a lower bound for $T_5$. From, (131),

$$\frac{dI(t)}{dt} \Big|_{t = \bar{T}_5} - \kappa I(\bar{T}_5) = -\beta \bar{N}_l(\bar{T}_3).$$

is satisfied. If $\bar{T}_5 \leq \bar{T}_3$, then $\bar{T}_5 \leq T_3 \leq T_5$. Suppose $\bar{T}_5 > \bar{T}_3$. Recall that $\bar{T}_3 \leq T_3 \leq T_5$ and $\bar{N}_l(\bar{T}_3) = N_l(\bar{T}_3)$ (from (132)). Then, $\bar{N}_l(\bar{T}_3) \leq N_l(T_5)$. Also, $\frac{dI(t)}{dt} - \kappa I(t)$ is a decreasing function of $t$ when its value is negative. Combining these facts with the definitions of $T_5$ and $\bar{T}_5$, we can assert that $\bar{T}_5 \leq T_5$.

*Case 2: The evolution of $N_l(t)$ and $N_i(t)$ takes place along Stage 4 and Stage 6.*

We have to consider two cases, $T_4 < \bar{T}_3$ and $T_4 \geq \bar{T}_3$ respectively.

Suppose $T_4 < \bar{T}_3$: First, we show that,

$$\bar{N}_l(\bar{T}_3) = N_l(\bar{T}_3). \tag{134}$$

Note that the dynamics of actual and the bounding processes are identical untill $t = T_4$. Then, $N_w(T_4) = \bar{N}_w(T_4)$. Also, during $T_4 < t \leq \min\{T_6, \bar{T}_3\}$, $\bar{N}_i(t)$ grows faster than $N_i(t)$, while $\bar{N}_l(t)$ grows at the same rate as that of $N_l(t)$. Therefore, to prove (134) holds true, we just need to show that $T_6 \geq \bar{T}_3$, which is done as follows: Note that, when $t \in [T_4, \min\{T_6, \bar{T}_3\}]$, the growth rate of $N_l(t) + N_i(t)$ is less than that of $\bar{N}_l(t) + \bar{N}_i(t)$, and hence $N_w(t) \leq N_w(t)$. Then, from (130) and the definition of $\bar{N}_w(t)$, we get $N_w(t) > 0$ when $T_4 < t < \bar{T}_3$ (since $T_4 > T_2$ by definition). Then, from (110), we get that $T_6$ cannot be less than $\bar{T}_3$.

Now, suppose $\bar{T}_5 \leq \bar{T}_3$. Then, from (134) and (127),

$$\bar{N}_l(T_\infty) = \bar{N}_l(\bar{T}_3) = N_l(\bar{T}_3) \leq N(T_\infty),$$

which proves our claim. Now, we show that $\bar{T}_5 \leq \bar{T}_3$ as follows: For all $t > T_4$, (109) is satisfied. Then, we get

$$\frac{dI(t)}{dt} \Big|_{t = \bar{T}_3} - \kappa I(\bar{T}_3) \leq -\beta N_l(\bar{T}_3).$$

due to the assumption, $T_4 < \bar{T}_3$ and the definition of $N_w(t)$. But, from (131) and (134),

$$\frac{dI(t)}{dt} \Big|_{t = \bar{T}_5} - \kappa I(\bar{T}_5) = -\beta N_l(\bar{T}_3).$$

Therefore, $\bar{T}_5 \leq \bar{T}_3$ since $\frac{dI}{dt} - \kappa I(t)$ is decreasing in $t$ once it goes negative.

Suppose $T_4 \geq \bar{T}_3$: Note that the dynamics of actual and the bounding processes are identical untill $t = \bar{T}_3$. To prove the claim, we show that

$$\frac{dN_l(t)}{dt} \geq \frac{d\bar{N}_l(t)}{dt} \quad \text{when} \quad t \geq \bar{T}_3. \tag{135}$$

At $t = \bar{T}_3$, from (129), the dynamics of actual and the bounding processes, the above expression holds true. Also, during $t \in [\bar{T}_3, T_6]$, $\frac{dN_l(t)}{dt}$ and $\frac{d\bar{N}_l(t)}{dt}$ are increasing and decreasing functions respectively. Hence, (135) holds true until $t \leq T_6$. Now, we show that $\bar{T}_5 < T_6$, and hence the growth rate of $\bar{N}_l(t)$ is zero for $t \geq T_6$. This asserts that (135) holds for $t \geq T_6$. The proof is as follows: From (99) and the definition of $N_w(t)$, we get

$$\frac{dI(t)}{dt} \Big|_{t = T_4} - \kappa I(T_4) = -\beta N_l(T_4) - (1 + \kappa) N_w(T_4). \tag{136}$$

Then, $\bar{T}_5 \leq T_4$ due to these reasons: 1) $\bar{T}_5$ satisfies (131), 2) $\beta \bar{N}_l(\bar{T}_3) = \beta N_l(\bar{T}_3 < T_4) + (1 + \kappa) N_w(T_4)$ since $\bar{T}_3 < T_4$ by assumption, 3) $\frac{dI(t)}{dt} - \kappa I(t)$ is decreasing once its value goes negative. Now, since $T_4 < T_6$, we have $\bar{T}_5 < T_6$, and hence (135) is attained.

Having shown that $\bar{N}_l(t)$ bounds $N_l(t)$ from below, we evaluate $\bar{N}_l(T_\infty)$ in the next step.

**Step 5:** *Evaluate the bounding process, $\bar{N}_l(T_\infty)$:*

*Find $\bar{N}_l(T_2)$:* The evolution of the bounding processes during $[0, T_2]$ are given by (127) and (128). Solving them, we get

$$\bar{N}_l(t) = (\Phi(I(t)) - \Phi(I(0))) e^{\beta t} + I(0) e^{\beta t},$$
$$= (\Phi(I(t)) + I(0)) e^{\beta t},$$

where $\Phi(x)$ is defined by (113). The second equality holds true since $\Phi(I(0)) = 0$.

Substituting $T_2$ from (118) in the above result,

$$\bar{N}_l(T_2) = (\Phi(I_2) + I(0)) e^{\beta T_2},$$

where $I_2 = I(T_2)$.

*Find $\bar{N}_l(\bar{T}_3)$:* Solving the growth equations given by (127) and (128), for the interval $[T_2, \bar{T}_3]$, we get

$$\bar{N}_l(t) = \left( \frac{C}{\beta} + \bar{N}_l(T_2) \right) e^{\beta(t - T_2)} - \frac{C}{\beta}.$$

Substituting, $\bar{T}_3$ from (122), and $\bar{N}_l(T_2)$ in the above expression, we get

$$N_l(\bar{T}_3) = \frac{C}{\beta} (e^{\beta \Delta T_2} - 1) + (\Phi(I_2) + I(0)) e^{\beta \bar{T}_3} = L_3.$$

where $L_3$ is given by (123).

*Let $\bar{T}_3 < \bar{T}_5$. Find $\bar{N}_l(\bar{T}_5)$:* Solving the growth equations given by (127) and (128), for the interval $[\bar{T}_3, \bar{T}_5]$, we get

$$\bar{N}_l(t) = (\Phi(I(t)) - \Phi(I(\bar{T}_3))) e^{\beta t} + \bar{N}_l(\bar{T}_3) e^{\beta(t - \bar{T}_3)},$$

Substituting $\bar{T}_3, \bar{T}_5$ and $\bar{N}_l(\bar{T}_3)$ in the above equation, we get

$$\bar{N}_l(t) = (\Phi(I_5) - \Phi(I_4)) e^{\beta t} + L_3 e^{\beta(\bar{T}_5 - \bar{T}_3)} = L_4,$$

where $I_5, I_4, L_3$ and $L_4$ are given by (124), (123), (123) and (126) respectively.

*Find $\bar{N}_l(T_\infty)$:* From (127), we have $\frac{d\bar{N}_l(t)}{dt} = 0$, for $t \geq \max\{\bar{T}_3, \bar{T}_5\}$. Therefore, we have $\bar{N}_l(T_\infty) = \bar{N}_l(\max\{\bar{T}_3, \bar{T}_5\})$. Then,

$$N_l(T_\infty) \geq \bar{N}_l(T_\infty) = \begin{cases} \bar{N}_l(\bar{T}_3) = L_3 & \text{if } \bar{T}_5 \leq \bar{T}_3 \\ \bar{N}_l(\bar{T}_5) = L_4, & \text{else.} \end{cases}$$

We have characterized the number of legal copies generated in the system in the presence of an efficient illicit P2P in the previous two lemmas. Attaining the statement in Theorem 3 is accomplished by studying the asymptotics of the results in Lemma 13 and 14. We start by introducing a few notation.

$$\Delta T_3 = \frac{1}{\kappa} \ln \left[ \kappa (1-\kappa) \ln N + (1-\kappa) \right],$$
$$\tilde{T}_3 = T_2 + \Delta T_3, \tag{137}$$
$$\Delta T_4 = \frac{1}{\kappa} \ln \left[ \frac{\kappa(1-\kappa)}{1+\kappa} \ln N + (1-\kappa) \right], \tag{138}$$
$$\tilde{T}_4 = T_2 + \Delta T_4. \tag{139}$$

Also, we say, $A_N \sim B_N$, if $\lim_{N \to \infty} \frac{A_N}{B_N} = 1$, $A_N \preceq B_N$, if $\lim_{N \to \infty} \frac{A_N}{B_N} \leq 1$. and, $A_N \succeq B_N$, if $\lim_{N \to \infty} \frac{A_N}{B_N} \geq 1$. Now, we are ready to prove the theorem.

As $N$ goes large, for any given $\kappa$, the assumption of Lemma 14 that $\kappa < 1 - \frac{2}{\sqrt{\ln N}}$ is attained. Therefore, in the asymptotic case, we use the result of Lemma 14. That lemma says,

$$N_l(T_\infty) \geq \begin{cases} L_3, & \text{if } \bar{T}_5 \leq \bar{T}_3 \\ L_4, & \text{else.} \end{cases} \tag{140}$$

where $\bar{T}_3, L_3, \bar{T}_5$ and $L_4$ are given by (122), (123), (125) and (126) respectively. The proof is done in two steps. First, we evaluate $L_3$. Next, we show that $\bar{T}_3 \succeq \bar{T}_5$. Then, from the above equation, we get that $N_l(T_\infty) \succeq L_3$.

*Evaluate $L_3$:* As $N$ goes larger, it can be shown that,

$$I_2 \sim \frac{N}{\ln N(1-\kappa)}, \quad \Delta T_2 \sim \frac{1}{\kappa} \ln \left( \kappa (1-\kappa) \log N \right),$$
$$T_2 \sim \ln \left( \frac{N}{I(0)(1-\kappa) \ln N} \right),$$
$$\bar{T}_3 \sim \ln \left[ \frac{N(\kappa(1-\kappa)\ln N)^{\frac{1}{\kappa}}}{I(0)(1-\kappa)\ln N} \right].$$
$$\Phi(I_2) \sim \left( \frac{I(0)}{N} \right)^\beta N \frac{(1-\kappa)}{(1-\beta)} \left( \frac{1}{(1-\kappa)\ln N} \right)^{1-\beta}.$$

The above results follows from (117), (121), (118), (122) and (113) respectively. Substituting the above results in (123), we get that

$$L_3 \sim \frac{N}{\ln N \beta} \left( \frac{(\ln N \kappa (1-\kappa))^{\frac{\beta}{\kappa}}}{(1-\beta)} - 1 \right). \tag{141}$$

*Show that $\bar{T}_3 \succeq \bar{T}_5$:* First of all, from (125) and (124), note that, $I(\bar{T}_5) = I_5$ and $I_5 \leq N$. Also, for large values of $N$, from (122) and the definition of $I(t)$, we can show that, $I(\bar{T}_3) \sim N$. Combining these two results, we get $I(\bar{T}_5) \preceq I(\bar{T}_3)$ This result in turn implies that $\bar{T}_5 \preceq \bar{T}_3$, since $I(t)$ is monotonically increasing.

Hence, from (140),

$$N_l(T_\infty) \succeq L_3.$$

From (141), the above equation, and (10), we get (16), which completes the first part of theorem.

The second part of the theorem deals with the case $\beta = 0$. From, (16), we have,

$$L \in \Omega \left( \frac{\ln \ln N}{\ln N} \right). \tag{142}$$

Now, to complete the proof, it suffices to prove the following lemma.

**Lemma 15.** *When $\beta = 0$,*

$$L \in o \left( \frac{\ln \ln N}{\ln N} \right)$$

.

*Proof:* Recall that when $\kappa < 1 - \frac{2}{\sqrt{\ln N}}$, which holds for any $\kappa$ when $N$ is large, the evolution of $N_l(t)$ and $N_i(t)$ takes place along the sequence of phases,'*Stage* 1 → *Stage* 2 →*Stage* 3 ( or *Stage* 4)→*Stage* 5 ( or Stage 6)'. We analyze each of these phases and obtain an upper bound on $N_l(T_\infty)$ as follows.

*Stage 1:* An upper bound on the number of legal copies at the end of this stage is given by,

$$N_l(T_2) \preceq \frac{N}{\ln N(1-\kappa)}. \tag{143}$$

which follows from the facts that $N_l(t) \leq I(t)$ for all $t$ and $I(T_2) \sim \frac{N}{\ln N(1-\kappa)}$. *Stage 2:* First we show that as $N$ goes large, $T_4 \preceq \bar{T}_3$ and hence, in the asymptotic case Stage 2 is followed by Stage 4. The proof of this claim proceeds as follows. Let, $U(t) = N_l(t) + N_i(t)$. From the dynamics of evolution of Stage 2, given by (93) and (94),

$$U(t) = \left( \frac{C}{\kappa} + I_2 \right) e^{\kappa(t-T_2)} - \frac{C}{\kappa}, \tag{144}$$

where $I_2$ is given (117) and $T_2$ is given by (118). Now, substituting $\tilde{T}_3$ from (137) in the above equation, we get

$$U(\tilde{T}_3) \sim I(\tilde{T}_3).$$

Also, it is easy to verify that $\tilde{T}_3$ satisfies (97). These results along with the definition of $T_3$, given by (97-98), implies that $\tilde{T}_3 \sim T_3$. Similarly, substituting $\tilde{T}_4$ in (144), we can show that

$$U(\tilde{T}_4) \sim \frac{1}{1+\kappa} \left( I(\tilde{T}_4) + \frac{dI}{dt}(\tilde{T}_4) \right).$$

This result along with the definition of $T_4$, given by (99), implies that $\tilde{T}_4 \sim T_4$.

We have, $\tilde{T}_4 \preceq \tilde{T}_3$, since

$$U(\tilde{T}_4) = \frac{N}{1+\kappa} < N = U(\tilde{T}_3),$$

and $U(t)$ is monotonically increasing. Therefore, we conclude that $T_4 \preceq T_3$. And hence, this stage is always followed by Stage 4.

Then, from the dynamics of $N_l(t)$, given by (94),

$$N_l(T_4) = N_l(T_2) + C_N(T_4 - T_2).$$

Now, from (143) and the definitions of $\tilde{T}_4$ and $T_2$, we get

$$N_l(T_4) \preceq \frac{N}{\ln N(1-\kappa)} + \frac{N}{\kappa \ln N} \ln \left( \ln N \frac{\kappa(1-\kappa)}{1+\kappa} + 1 - \kappa \right).$$
(145)

*Stage 4:* This stage starts at time $t = T_4$. From the discussion given above (in Stage 3 analysis), $T_4 \sim \tilde{T}_4$. Then, from (139), $I(T_4) \sim I(\tilde{T}_4) \sim N$ and $\frac{dI}{dt}(T_4) \sim \frac{dI}{dt}(\tilde{T}_4) \sim 0$. Also, $N_w(T_4) = I(\tilde{T}_4) - U(\tilde{T}_4) \sim \frac{N\kappa}{1+\kappa}$. Recall that $U(t) = N_l(t) + N_i(t)$. And $U(\tilde{T}_4)$ is obtained from (144) and (139).

Using these facts and the dynamics of $N_i(t)$ and $N_l(t)$ given by (108) and (107) respectively, we show that,

$$U(t) = (C_N + N)(1 - e^{-t}) + U(\tilde{T}_4)e^{-(t-\tilde{T}_4)}.$$

This stage terminates, when no Wanters are left to be served, i.e $U(t) \sim N$. Let $\tilde{T}_6$ marks this event. Then,

$$\tilde{T}_6 \sim \ln \left( \frac{\ln N}{1 + \kappa} \right).$$

The legal copies of content generated in this phase is $C_N \times (\tilde{T}_6 - \tilde{T}_4)$ from the dynamics of $N_l(t)$ given by (107). Then, from the above result and (145), we get

$$N_l(T_\infty) \preceq \frac{N}{\ln N} \ln \left[ \frac{(\ln N)^{(\frac{1}{\kappa}+1)}}{1+\kappa} \left( \frac{(1-\kappa)\kappa}{(1+\kappa)} \right)^{\frac{1}{\kappa}} \right],$$

which completes the proof. ∎

## APPENDIX E
## TECHNICAL LEMMAS

**Lemma 16.** *Consider a differential equation given*

$$\frac{dy}{dt} = C_N + \frac{\beta y}{N}(I - U(t))$$
(146)

*where*

$$U(t) = \frac{N\theta_2}{\kappa} + \frac{N\Delta\theta/\kappa}{1 + be^{-\Delta\theta(t-T)}}.$$

*Then for all $t - T > \frac{\ln b}{\Delta\theta}$, the solution to the above differential equation satisfies the inequality*

$$y(t) \geq y(T) \left( \frac{1+b}{d} \right)^{\frac{\beta}{\kappa}} e^{(-q_1(t-T))}$$

$$+ C_N \left( \frac{b}{d} \right)^{\frac{\beta}{\kappa}} e^{(-q_1(t-T))} \left( \frac{e^{\left(q_1 \frac{\ln b}{\Delta\theta}\right)}}{q_1} - \frac{1}{q_1} \right) \mathbf{1}_{b \geq 1}$$

$$+ C_N \left( \frac{1}{d} \right)^{\frac{\beta}{\kappa}} e^{(-q_1(t-T))} \left( \frac{e^{(q_2 \Delta \tau_j)}}{q_2} - \frac{e^{\left(q_2 \frac{\ln b}{\Delta\theta}\right)}}{q_2} \mathbf{1}_{b \geq 1} \right)$$

$$- C_N \left( \frac{1}{d} \right)^{\frac{\beta}{\kappa}} e^{(-q_1(t-T))} \frac{1}{q_2} (1 - \mathbf{1}_{b \geq 1}),$$
(147)

*where $d = (b + \exp(\Delta\theta(t - T)))$, $q_1 = \left( \frac{\beta\theta_2}{\kappa} - \frac{\beta I}{N} \right)$ and $q_2 = \frac{\beta\theta_1}{\kappa} - \frac{\beta I}{N}$. Furthermore, for $\beta = 0$, equality holds.*

*Proof:* A general solution to the above differential equation is

$$y(t) = \frac{\int C_N \exp(\int P dt) + M}{\int P dt}$$
(148)

where $P(t) = -\frac{\beta}{N}(I - U(t))$. We have

$$\int P dt = -\frac{\beta I t}{N} + \frac{\beta\theta_2 t}{\kappa} + \frac{\beta}{\kappa} \ln(1 + (1/b)\exp(\Delta\theta(t - T))).$$

Then,

$$C_N e^{\int P dt} = C_N B(t) \exp\left( \frac{\beta\theta_2}{\kappa} - \frac{\beta I t}{N} \right) t,$$

where

$$B(t) = (1 + (1/b)\exp(\Delta\theta(t - T)))^{\frac{\beta}{\kappa}}.$$

For $b \geq 1$, we can lower bound $B(t)$ as

$$B(t) \geq \begin{cases} 1 & t \leq \frac{\ln b}{\Delta\theta} + T \\ \left( \frac{1}{b} \right)^{\frac{\beta}{\kappa}} \exp\left( \frac{\beta}{\kappa} \Delta\theta(t - T) \right) & t > \frac{\ln b}{\Delta\theta} + T. \end{cases}$$
(149)

On the other hand, if $b < 1$,

$$B(t) \geq \left( \frac{1}{b} \right)^{\frac{\beta}{\kappa}} \exp\left( \frac{\beta}{\kappa} \Delta\theta(t - T) \right), \quad \forall t.$$
(150)

Let us now evaluate $A(t)$. We have

$$A(t) = \int C_N e^{\int P dt} dt.$$

Initially consider the case $b \geq 1$. For $t < \frac{\ln b}{\Delta\theta} + T$, it is easy to verify that

$$A(t) \geq C_N \frac{\exp\left( \left( \frac{\beta\theta_2}{\kappa} - \frac{\beta I}{N} \right) t \right)}{\frac{\beta\theta_2}{\kappa} - \frac{\beta I}{N}}$$
(151)

where the inequality follows from (149). For $t > \frac{\ln b}{\Delta\theta} + T$, we have

$$A(t) \geq A(\frac{\ln b}{\Delta\theta} + T) + \int_{\frac{\ln b}{\Delta\theta} + T}^{t} C_N e^{\int P dt}$$
(152)

$$\geq C_N \exp(q_1 T) \exp\left( q_1 \frac{\ln b}{\Delta\theta} \right) \frac{1}{q_1}$$

$$+ C_N \exp(q_1 t) \left( \frac{1}{b} \right)^{\frac{\beta}{\kappa}} \frac{\exp\left( \frac{\beta\Delta\theta}{\kappa}(t - T) \right)}{q_2}$$

$$- C_N \exp(q_1 T) \left( \frac{1}{b} \right)^{\frac{\beta}{\kappa}} \frac{\exp\left( q_2 \frac{\ln b}{\Delta\theta} \right)}{q_2}.$$

where $q_1 = \left( \frac{\beta\theta_2}{\kappa} - \frac{\beta I}{N} \right)$ and $q_2 = \frac{\beta\theta_1}{\kappa} - \frac{\beta I}{N}$.

In the second case, in which $b < 1$, for all values of $t$, we have,

$$A(t) \geq C_N \exp(q_1 t) \left( \frac{1}{b} \right)^{\frac{\beta}{\kappa}} \frac{\exp\left( \frac{\beta\Delta\theta}{\kappa}(t - T) \right)}{q_2}.$$

where the inequality follows from (150).

Then, combining the expressions of $A(t)$ in both cases, for $t > \frac{\ln b}{\Delta\theta} + T$, we have,

$$A(t) \geq C_N \exp(q_1 T) \exp\left( q_1 \frac{\ln b}{\Delta\theta} \right) \frac{1}{q_1} \mathbf{1}_{b \geq 1}$$
(153)

$$+ C_N \exp(q_1 t) \left( \frac{1}{b} \right)^{\frac{\beta}{\kappa}} \frac{\exp\left( \frac{\beta\Delta\theta}{\kappa}(t - T) \right)}{q_2}$$

$$- C_N \exp(q_1 T) \left( \frac{1}{b} \right)^{\frac{\beta}{\kappa}} \frac{\exp\left( q_2 \frac{\ln b}{\Delta\theta} \right)}{q_2} \mathbf{1}_{b \geq 1}.$$

where $\mathbf{1}_{b \geq 1}$ is the indicator function defined by (51).

Using the above result in equation (148), we get that for $t > \frac{\ln b}{\Delta \theta} + T$,

$$y(t) = \frac{M}{\exp(\int P dt)} + \frac{A(t)}{\exp(\int P dt)} \tag{154}$$

$$\begin{aligned}
&\geq M \left(\frac{b}{d}\right)^{\frac{\beta}{\kappa}} \exp\left(-q_1 t\right) \\
&+ C_N \left(\frac{b}{d}\right)^{\frac{\beta}{\kappa}} \exp\left(-q_1(t-T)\right) \exp\left(q_1 \frac{\ln b}{\Delta \theta}\right) \frac{1}{q_1} \mathbf{1}_{b \geq 1} \\
&+ C_N \left(\frac{1}{d}\right)^{\frac{\beta}{\kappa}} \frac{\exp\left(\frac{\beta \Delta \theta}{\kappa}(t-T)\right)}{q_2} \\
&- C_N \left(\frac{1}{d}\right)^{\frac{\beta}{\kappa}} \exp\left(-q_1(t-T)\right) \frac{\exp\left(q_2 \frac{\ln b}{\Delta \theta}\right)}{q_2} \mathbf{1}_{b \geq 1}.
\end{aligned} \tag{155}$$

where $d = (b + \exp(\Delta\theta(t-T)))$.

Using boundary conditions, we can show that

$$\begin{aligned}
M &= \left(\frac{1+b}{b}\right)^{\frac{\beta}{\kappa}} \exp\left(q_1 T\right) \left(y(T) - C_N \left(\frac{b}{1+b}\right)^{\frac{\beta}{\kappa}} \frac{1}{q_1} \mathbf{1}_{b \geq 1}\right) \\
&- \left(\frac{1+b}{b}\right)^{\frac{\beta}{\kappa}} \left(C_N \left(\frac{1}{1+b}\right)^{\frac{\beta}{\kappa}} \frac{1}{q_2}(1 - \mathbf{1}_{b \geq 1})\right).
\end{aligned}$$

Substituting the above equation in equation (155) and rearranging yields (147). For $\beta = 0$, the inequalities in equations (149) and (150) become equalities and we get the lemma. ∎